\documentclass[12pt,a4paper]{article}
\usepackage{amsfonts}
\usepackage{amsmath}
\usepackage{bm}
\usepackage{fancyhdr}
\usepackage[a4paper, total={6.5in, 10in}]{geometry}
\usepackage{amssymb}
\usepackage{float}
\usepackage{caption}
\usepackage[caption=false]{subfig}
\usepackage{graphicx}
\usepackage{booktabs}
\usepackage{chngcntr}
\usepackage{placeins}
\usepackage{authblk}%
\usepackage{rotating}
\usepackage{tikz}
\usetikzlibrary{positioning}
\usepackage[colorlinks=true, citecolor=blue, linkcolor=blue, urlcolor=blue]{hyperref}
\usepackage[natbibapa]{apacite}
\providecommand{\U}[1]{\protect\rule{.1in}{.1in}}

\newtheorem{theorem}{Theorem}

\newtheorem{proposition}[theorem]{Proposition}

\newenvironment{proof}[1][Proof]{\noindent\textbf{#1.} }{\ \rule{0.5em}{0.5em}}

\hypersetup{
  colorlinks   = true,
  urlcolor     = blue,
  linkcolor    = blue,
  anchorcolor = blue
  citecolor   = blue
}

\begin{document}

\title{How many parents does it take? Parental time allocation and the effectiveness of fertility subsidies\thanks{This paper will be presented at the 101st WEAI Annual Conference, Denver, United States; and at the ASSA 2027 Annual Meeting, Washington DC, United States. Special thanks are due to Daniele Tavani, Elissa Braunstein, and Ray Miller for stimulating discussions during the preparation of this article. The usual caveats apply.}}
\author{Jackie Dajin Young$^{\dagger}$ \: $\cdot$ \: Marwil J. Dávila-Fernández$^{\dagger}$}

\date{{\normalsize {$^{\dagger}$\emph{Colorado State University}}} \bigskip \bigskip \\
May 2026} 

\maketitle

\begin{abstract}
There has long been an apparent consensus in the literature on intra-household allocation and fertility that greater paternal involvement in childcare relaxes maternal time constraints, enabling mothers to increase their labor supply or leisure. Recent evidence, particularly from South Korea, challenges this view: increases in fathers' childcare time have coincided with a further increase in mothers' time dedicated to child-rearing. This paper develops an Overlapping Generations (OLG) growth model to address such a puzzle. The central mechanism and our main innovation hinge on the functional form of the \textit{childcare technology}. When maternal and paternal time are substitutes, the conventional result holds. However, when they are complements, greater paternal involvement necessarily raises maternal childcare time, depressing fertility and redirecting household resources toward child quality. We further argue that the elasticity of substitution should not be interpreted as a pure preference parameter, as it also reflects the social and institutional norms, the skills each parent brings to child-rearing and their intergenerational transmission. The model is extended to study the effectiveness of pro-natalist subsidies, suggesting that such policies may generate an unintended anti-fertility bias. Numerical simulations calibrated loosely to South Korean data confirm that the model is consistent with the observed quantity-quality trade-off and the persistence of low fertility despite active pro-natalist policy.

\bigskip

  \bigskip\ 


\end{abstract}

\newpage

\section{Introduction}

A recurring finding in the literature on intra-household time allocation is that the division of childcare between mothers and fathers is unequal, with mothers bearing the larger share. A common policy intuition in low-fertility settings is that greater paternal involvement can relax mothers' care constraints, lowering the effective cost of childbearing and, at least potentially, supporting higher fertility (e.g. \citealp{Tamm2019}; \citealp{ThomasEtAl2022LeavePolicies}; \citealp{PerssonRossinSlater2024}). This logic is intuitive and has informed family policy in a number of countries facing sustained fertility decline. Yet feminist economics and gender-demographic approaches have long emphasized that additional father-child time need not mechanically displace mothers' responsibility for supervision, planning, and default accountability \citep{Folbre2001, McDonald2000, GoldscheiderBernhardtLappegard2015}.

South Korea offers a striking case through which to examine this premise. Over the past several decades, the country has experienced one of the most dramatic fertility declines ever recorded, reaching a total fertility rate of 0.75 in 2024, while paternal involvement in childcare has also increased. Using comparable Korean Time Use Survey (KTUS) waves, \citet{Park2021} shows that fathers' childcare time more than doubled between 1999 and 2014. Yet mothers' childcare time did not fall over the same period. It rose as well, from approximately 130 to more than 200 minutes per day. Although the KTUS underwent changes after 2014 that complicate direct comparisons with later waves, official 2019 estimates continue to show a large maternal care burden. Fathers are doing more than before, but mothers are not doing less. This pattern does not prove complementarity between parental time inputs, but it is difficult to reconcile with a strong substitution view in which additional father time mechanically displaces mother time. It sits alongside a dramatic rise in educational attainment and persistently high private education expenditure. Falling fertility, rising parental time inputs, and increasing child-quality expenditure driven in part by positional educational competition are difficult to square with models in which paternal and maternal time in childcare are treated as freely interchangeable.

The present paper develops a novel Overlapping Generations (OLG) growth model designed to account for this apparent puzzle. The central innovation is the treatment of the childcare technology. Rather than assuming that parental time inputs are freely substitutable, we model effective childcare time as a CES aggregator of maternal and paternal hours. At the household level, we are able to vary the degree of substitutability between parental inputs and trace its implications for fertility and education investment. At the macro level, we trace the consequences of those optimal allocations for the aggregate dynamics of human capital and population. The model delivers two sets of results. The first concerns the positive implications of the childcare technology. When maternal and paternal time are substitutes, greater paternal involvement eases the maternal time burden, reducing the effective cost of children.

However, when parental time inputs are complements, the opposite occurs. An increase in paternal childcare time necessarily raises maternal childcare time as well, tightening the household time constraint and pushing fertility down. Households respond by reducing the number of children and investing more heavily in each child's education, exacerbating the quantity-quality trade-off. Still, at the aggregate level, while lower fertility results in smaller populations, decreasing returns to education ensure that human capital remains the same across regimes. We argue that the elasticity of substitution between parental time inputs should not be interpreted as a pure preference parameter. It reflects the skills that mothers and fathers bring to child-rearing, which are shaped by gender norms, social expectations, and the intergenerational transmission of household competencies. A father who can perform the same tasks as a mother is closer to a substitute; one who cannot is closer to a complement.

The second set of results concerns the effectiveness of pro-natalist subsidies. We consider two fiscal arrangements. In the first, the government runs a balanced budget at all times and households internalize the implied tax burden, embedding a form of Ricardian equivalence by design. In the second, the tax rate that balances the government's budget is calculated ex post, after households have made their fertility and education decisions. Under this arrangement, fertility subsidies have a distortionary effect on household choices. We show that under complementarity, pro-natalist subsidies may generate an unintended anti-fertility bias. The model admits two optimal allocations of education and fertility. In one of them, the subsidy raises the cost of the tax required to finance it in a way that more than offsets the direct incentive to have children. This mechanism provides one possible explanation for why cash-based pro-natalist policies may have limited effects when they do not alter the underlying organization of care and child-investment expectations.

The paper contributes to a broad tradition of OLG models applied to demographic and macroeconomic questions. Their structure has been used to study inflation and savings dynamics \citep{BernasconiKirchkamp2000, SterkTenreyro2018}, optimal monetary policy rules \citep{CrettezEtAl2002, vonThadden2012, Hiraguchi2014}, rational asset price bubbles \citep{Gali2014}, the non-neutrality of monetary policy \citep{BraunIkeda2021, HuEtAl2023}, and environmental issues \citep{JohnPecchenino1994, CaravaggioSodini2023, DavilaFernandezEtAl2025}. The structure of our model follows \citet{delacroix_gosseries_2012}, abstracting from the environmental component \citep[see also][]{delacroix2002theory}. Within the strand of gender OLG models that study intra-household time allocation, the closest references are perhaps Agénor and coauthors. An important innovation of our framework is that we do not impose gender-based specialization in child-rearing. Much of the existing literature assumes that mothers bear the entirety or the majority of the time cost associated with raising children, effectively specializing in this activity within the household \citep{Agenor2017computable}. We depart from this assumption, allowing both parents to contribute to child-rearing without exogenous gender-based constraints, and show that the degree of substitutability between their time inputs is central to understanding both fertility outcomes and the effectiveness of pro-natalist policy \citep[see also][]{Agenor2021GenderGaps, AgenorAg2023Infrastructure}.

The remainder of the paper is organized as follows. Section 2 reviews the relevant literature and documents the stylized facts from South Korea that motivate the analysis. Section 3 develops the baseline model and analyzes the two polar cases of perfect substitutability and perfect complementarity, differentiating between two different treatments of the government budget. Section 4 extends the model to study distortionary fertility subsidies. Section 5 concludes with some final considerations and avenues for future research.

\section{Motivation and literature review}

A large literature in household economics has studied how parents allocate time within the family and whether changes in that allocation can alter mothers' constraints and fertility behavior. In a seminal contribution, \citet{Becker1965} models time allocation through household production and specialization, with spouses dividing market and non-market work according to their relative productivities. The bargaining literature that emerged in the 1980s and 1990s constitutes, in part, a critique of this unitary framework. Early formulations by \citet{ManserBrown1980} and \citet{McElroyHorney1981} model household decisions as the outcome of negotiation between individuals with distinct preferences. \citet{LundbergPollak1993} and \citet{LundbergPollak1996} further emphasize the role of separate spheres, noncooperative threat points, and the institutional structure of marriage. Subsequent work has extended these foundations into collective, bargaining, and intertemporal models of household behavior (see \citealp{ChiapporiMazzocco2017} for a comprehensive review).

While the present paper is informed by that tradition, our primary goal is not to treat household bargaining or the division of domestic labor in full generality. Our focus is narrower on the economics of \emph{childcare}, more specifically, whether greater paternal involvement meaningfully reduces the effective time burden that children place on mothers. This distinction matters because childcare differs from broader domestic labor in several economically important ways. It is more time-sensitive, less easily postponed, and more tightly bound to norms and skills related to parental responsibility. For instance, although a parent can defer housework, childcare, especially for young children, is far less deferrable. Moreover, an observed increase in fathers' time with children need not imply a comparable reduction in mothers' burden if mothers retain primary responsibility for supervision, planning, cognitive labor, and default accountability \citep{Craig2006, Daminger2019}. For that reason, evidence from the broader domestic-labor literature does not map cleanly into the question we study here. Our interest is not in whether household labor becomes more equal in some broad sense, but in whether mothers' and fathers' \emph{care inputs} are substitutable in a way that lowers the effective cost of childbearing.

\subsection{A closer look at the care literature}

Using U.S. time-diary data, \citet{Bianchi2000} 
and \citet{SayerBianchiRobinson2004} document that both mothers' and fathers' time with children increased over time, even as maternal labor force participation rose. \citet{BianchiRobinsonMilkie2006} reach a similar conclusion. Fathers increased their childcare time from a low base, while mothers remained the primary providers of unpaid care. Crucially, the growth in paternal involvement has not translated into one-for-one reductions in maternal unpaid labor. The historical pattern that emerges is one in which fathers do more than before, but mothers continue to bear a large share of the care burden.

Paternal participation and paternal responsibility are not the same thing. \citet{RaleyBianchiWang2012} show that mothers are more likely to be relieved when fathers take on solo or routine care responsibilities rather than simply spending more time with children. This distinction is crucial for interpreting aggregate trends. Time-use data may show more father-child time without establishing whether fathers have displaced maternal effort in the dimensions of care that matter most for mothers' daily constraints. In that sense, the relevant margin is not simply paternal presence, but whether fathers absorb tasks that would otherwise remain mothers' responsibility.

Evidence from policy settings points in a similar direction. A number of studies have exploited paternal leave reforms as quasi-natural experiments in fathers' care involvement. For instance, \citet{Patnaik2019} shows that reserving parental leave exclusively for fathers increases take-up and has persistent effects on paternal involvement in childcare. Studying Germany, \citet{Tamm2019} finds that fathers' leave-taking raises their time in childcare and housework, providing evidence of some substitution between parental time inputs. More recently, \citet{PerssonRossinSlater2024} show that fathers' workplace flexibility around childbirth improves maternal postpartum health, suggesting that paternal temporal availability generates meaningful household spillovers even when the mechanism is not directly observed in time-use data. These studies confirm that paternal involvement can matter for mothers' constraints. However, they do not support a strong substitution view in which additional father time fully displaces the maternal burden. A more careful reading is one of partial relief, conditional on the type of care fathers provide and the institutional setting in which families make decisions.

Recent work also suggests that policy design alone does not guarantee large changes in the division of care. \citet{JorgensenSogaard2024} show, using Danish administrative data, that parents respond strongly to parental-leave incentives but also display a willingness to pay for gender-traditional leave allocations, with fathers taking little or no leave. Similarly, \citet{Rosenqvist2024} finds that a Swedish cash bonus rewarding a more equal division of leave had only modest effects on within-couple leave differences and no average effect on later earnings gaps or childcare responsibility. At a broader level, \citet{KlevenEtAl2024} observes that the large long-run expansion of parental leave and childcare policies in Austria had virtually no impact on gender convergence in earnings. These recent contributions temper a simplistic policy optimism. Family policy can affect behavior on some margins, but deeper gender asymmetries in care are often remarkably persistent (see also \citealp{ThomasEtAl2022LeavePolicies}).

The literature on bargaining and gender norms provides additional elements on why such persistence arises. \citet{BittmanEtAl2003} argue that gender trumps money. Women reduce their housework as their relative earnings rise, but only up to a point, beyond which normative expectations constrain further equalization. \citet{BertrandKamenicaPan2015} provide complementary evidence that gender identity norms shape household behavior when wives approach or exceed their husbands in earnings. More directly relevant to childcare, \citeauthor{IchinoOlssonPetrongoloSkogmanThoursie2019} 
(\citeyear{IchinoOlssonPetrongoloSkogmanThoursie2019}, \citeyear{IchinoOlssonPetrongoloSkogmanThoursie2024}) argue that the degree of substitutability between maternal and paternal childcare time depends on prevailing gender norms. Changes in relative prices or policy incentives need not produce large reallocations of care if couples remain attached to a norm of maternal primacy. This literature is especially relevant for our framework because it suggests that the effective technology of care within the household is not purely technical in nature. It is socially mediated, and the parameter governing substitutability in our model can be read as capturing that social dimension.

Feminist economics makes a related point from a different angle. 
\citet{Folbre2001} emphasizes that care is not well understood if treated as a standard household commodity because it is bound up with responsibility, supervision, and social obligation in ways that market goods are not. \citet{Braunstein2015} and \citet{BraunsteinVanStaverenTavani2011} further stress that unpaid care is central to social reproduction and remains unequally allocated across gender. These contributions do not provide a direct model of parental time substitution. Rather, they motivate why observed hours may be an incomplete measure of the true care burden. Care is embedded in responsibility, coordination, and social reproduction, so the same measured hour may carry different implications depending on who remains the default accountable parent. Even as fathers' observed time with children rises, mothers may continue to bear a disproportionate burden if they remain responsible for ensuring that care is actually delivered (for OLG examples using a related argument, see \citealp{Agenor2017computable}; \citealp{AgenorAg2023Infrastructure}).

\subsection{South Korea as a case example}

\begin{figure}[tbp]
    \centering
    \subfloat[]{\includegraphics[width=3.125in]{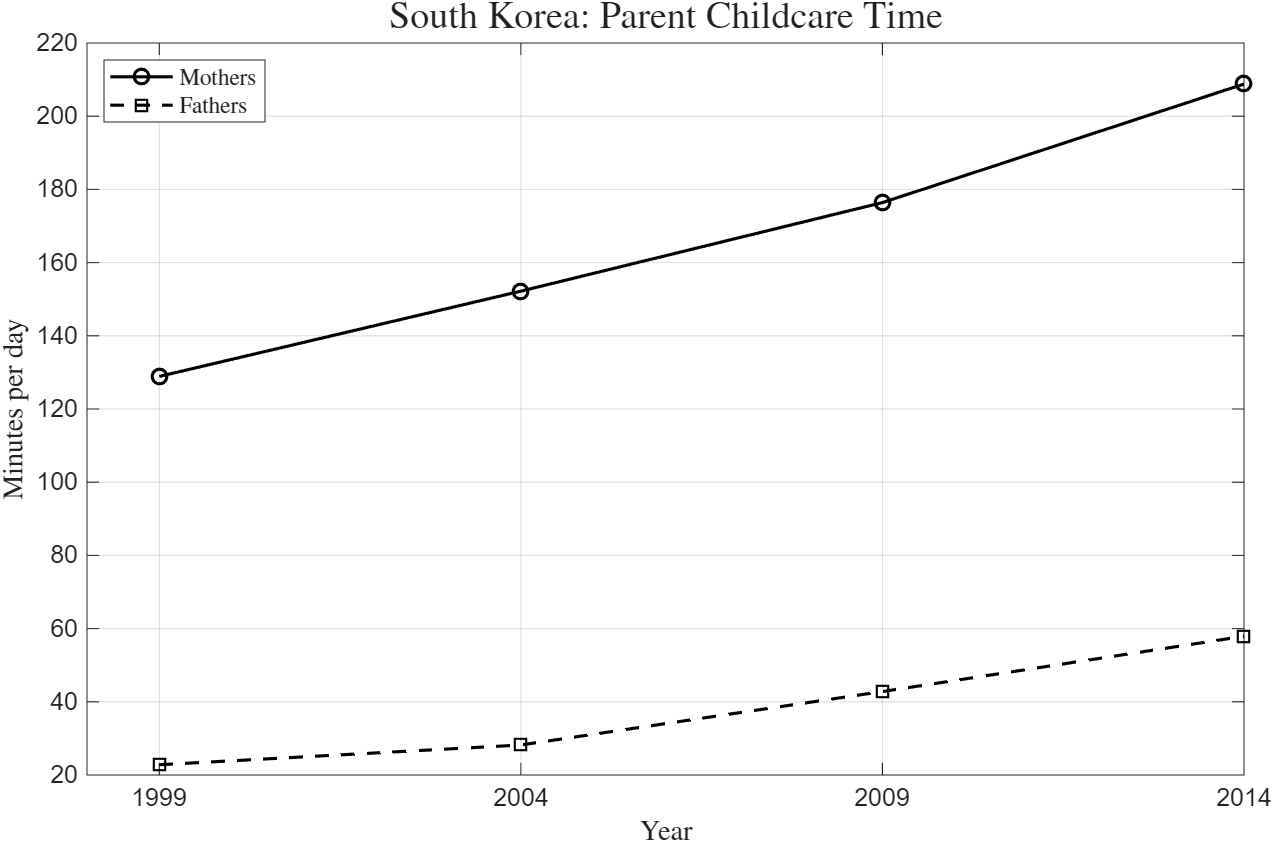}} \quad
    \subfloat[]{\includegraphics[width=3.1in]{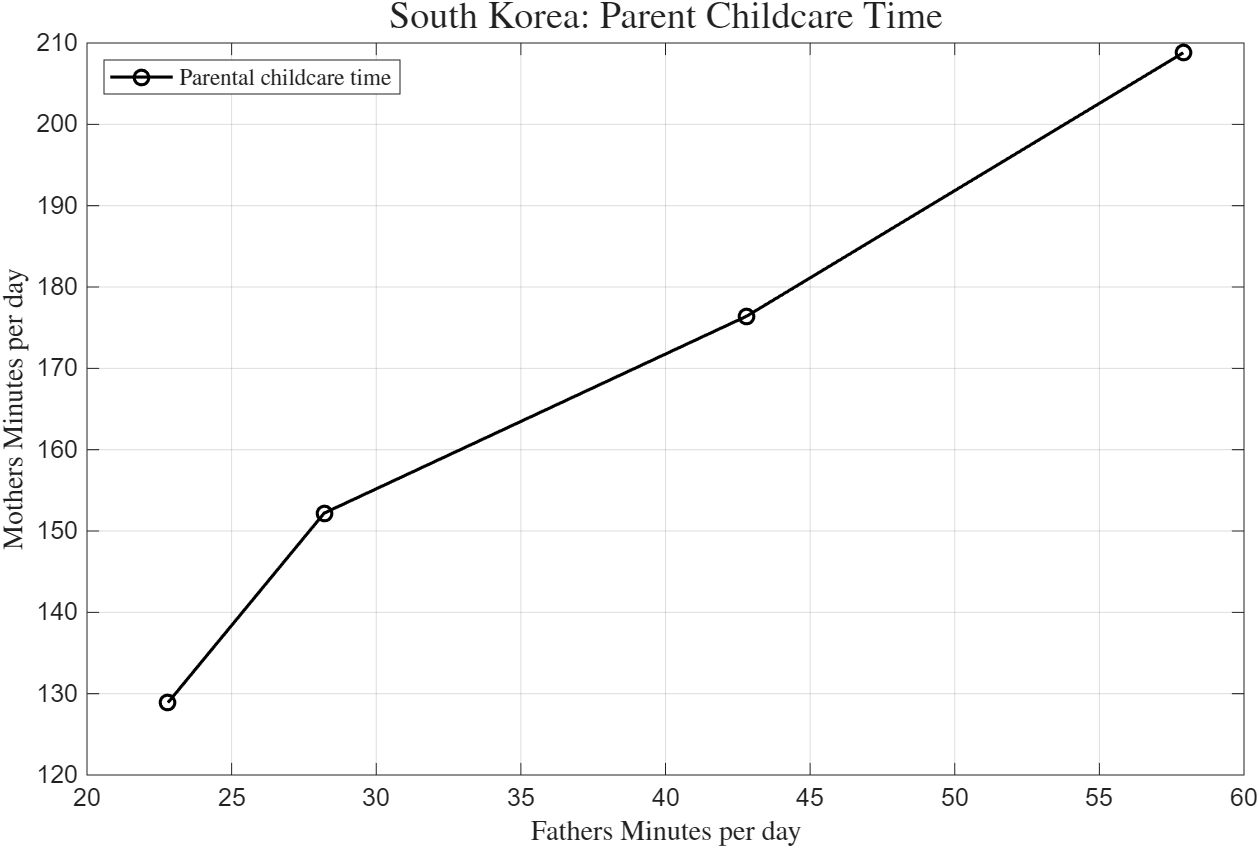}}
    \caption{Mothers' and fathers' childcare time in South Korea. 
    Source: \citet{Park2021} using Korean Time Use Survey (KTUS) 
    estimates for married parents with at least one child under school 
    age.}
    \label{fig:korea_care}
\end{figure}

With these considerations in mind, we turn to the Korean case. Fig.~\ref{fig:korea_care} indicates that while fathers' childcare time more than doubled between 1999 and 2014, rising from less than 25 minutes per day to around 60 minutes, mothers' childcare time did not fall over the same period. It actually rose from approximately 130 to more than 200 minutes per day. Fathers are doing more than before, but mothers are not doing less.

Such trajectories are not merely an artifact of the pre-2014 sample. Official 2019 time-use summaries point in the same direction. Among parents of children under age 10, mothers devoted 3 hours and 13 minutes per day to childcare, compared with 1 hour and 2 minutes for fathers.\footnote{The \textit{Korean Time Use Survey} revised its activity classification after 2014. The 2019 revision incorporated ICATUS 2016, refined caregiving categories, and changed the definition of ``presence of others,'' which is especially relevant for measuring passive or supervisory care. These changes complicate direct comparison with earlier waves. For this reason, Fig.~\ref{fig:korea_care} is limited to the comparable 1999--2014 waves used by \citet{Park2021}. The 2019 survey nonetheless confirms that the broader pattern remains: mothers of children under age 10 continued to devote substantially more time to childcare than fathers.} In dual-earner households, the gap in unpaid work is similarly pronounced: wives spent more than three times as much time on unpaid work as husbands. Taken together, the evidence does not establish complementarity by itself, but it is difficult to reconcile with a strong-substitution interpretation in which additional father time mechanically displaces mother time. If fathers' time displaced mothers' time one-for-one, rising paternal involvement should reduce mothers' childcare time or at least hold total parental childcare time roughly constant. In the comparable 1999--2014 series, however, total parental childcare time rose, and the absolute gap between mothers' and fathers' childcare time widened.

\begin{figure}[tbp]
    \centering
    \subfloat[]{\includegraphics[width=3.125in]{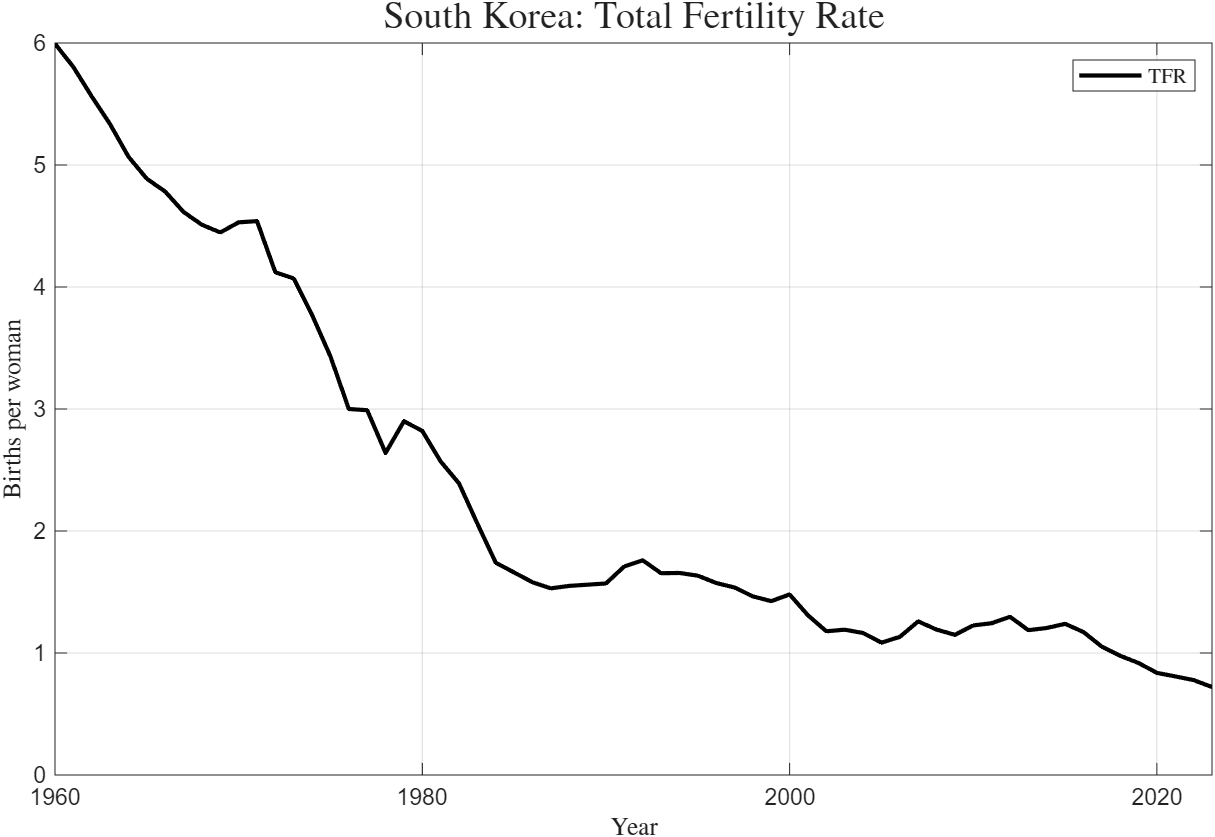}} \quad
    \subfloat[]{\includegraphics[width=3.125in]{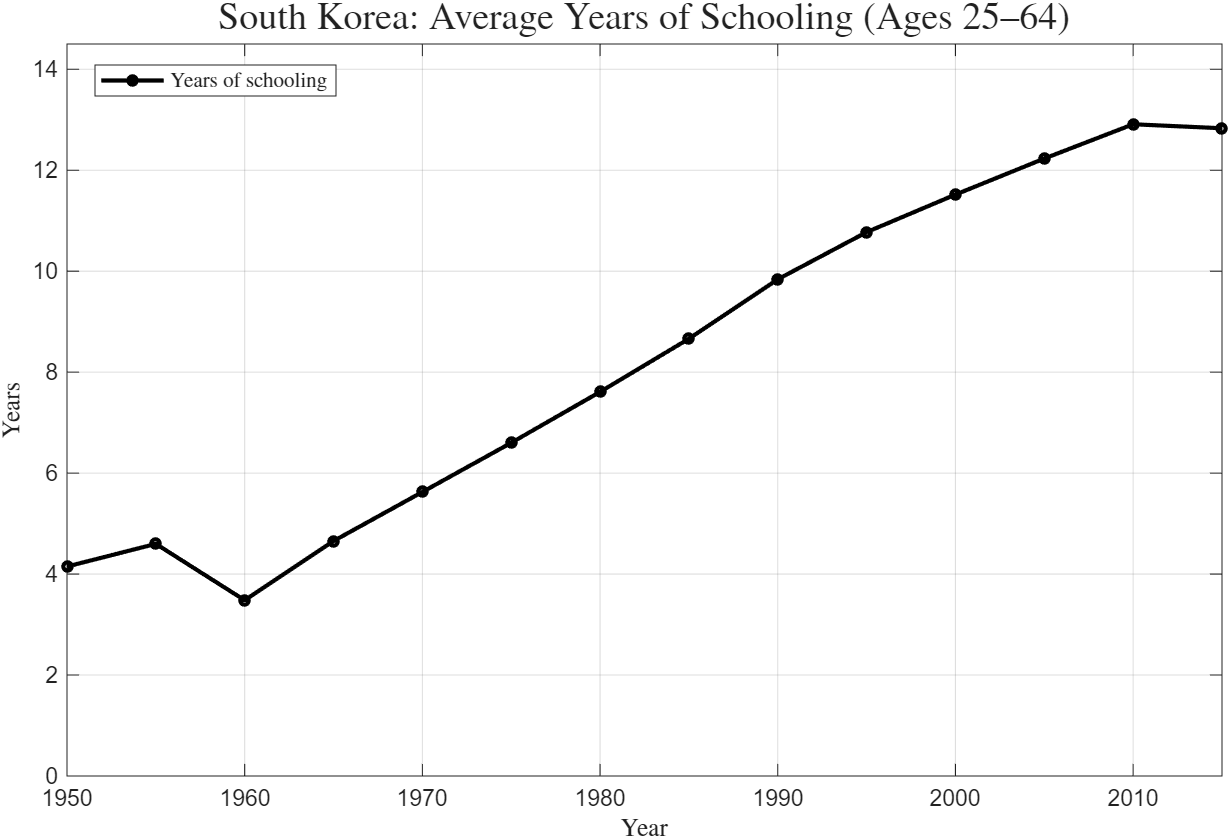}}
    \caption{Total fertility rates and average years of schooling in South Korea. Panel (a) covers the period 1960--2024 and uses data from the World Bank annual series through 2023 and 2024 official birth statistics from Statistics Korea. Panel (b) relies on Barro-Lee Educational Attainment Data.}
    \label{fig:korea_fertility}
\end{figure}

The apparent time-allocation puzzle is embedded in a broader set of transformations. Over the past six decades, Korea has experienced one of the most dramatic fertility declines ever recorded. Fig.~\ref{fig:korea_fertility} (a) shows that the total fertility rate fell from 6 in the 1960s to below 2 in the 1980s, reaching 0.75 in 2024. Over the same period, panel (b) reports that educational attainment rose sharply, with average years of schooling jumping from 4 to 13. These two trends are connected. As schooling expanded and returns to education remained high, the standard for what it means to raise a child well shifted upward. Parents are not simply deciding whether they can afford another child. They are deciding whether they can provide enough time, attention, and investment for another child to meet increasingly demanding expectations. Such pressures are visible in current private education spending. In 2024, total private education expenditures reached 29.2 trillion won, with a participation rate of 80.0 per cent and average monthly spending of 474,000 won per student \citep{MinistryDataStats2025PrivateEducation}. In this context, private education spending should be understood not only as investment in human capital, but also as part of a positional competition in which private incentives may exceed social returns \citep{AndersonKohler2013, KimTertiltYum2024}.

The policy response to these trends has centered on two main instruments. First, governments have promoted paternal involvement in childcare, premised on the idea that redistributing the care burden within the household will free up time for mothers and make additional children more feasible. Second, they have expanded fertility subsidies and other transfers designed to reduce the financial cost of having children. Neither instrument, on its own, appears sufficient to address the organization of care and child-investment expectations that shape the effective cost of fertility.

\subsubsection{South Korea's related literature}

Three strands of literature speak directly to the patterns documented above. The first looks specifically at parental time allocation. \citet{Park2021} documents that both mothers' and fathers' childcare time increased between 1999 and 2014 in South Korea, although mothers remained the primary caregivers throughout. His key point is not simply that fathers are doing more than before. It is that their increased involvement has not resulted in a reduction in mothers' childcare time in the aggregate trend. That is the pattern the model is trying to speak to.

The second asks whether greater support from husbands affects fertility intentions or actual births. Several papers suggest that it does. \cite{ParkChoChoi2010} find that paternal investment in childcare and housework is positively associated with women's intention to have a second child, especially among employed women. Along similar lines, \citet{Yoon2017} also finds that husbands' support for housework and childcare increases the likelihood of a second birth, but the result is not linear. In her estimates, the effect follows a threshold pattern. Only the highest level of husbands' daily support is clearly associated with second births. \citet{Kim2017DomesticLabour} obtains a similar pattern for intended second births. Women whose husbands contribute more domestic labor are more likely to have a second child, and the association is stronger when the husband's contribution is larger.

While this literature is useful, one should approach it carefully. Those studies do not imply that any increase in fathers' time will materially reduce the cost of another child. If anything, some of the evidence suggests the opposite. Support seems to matter when it is large enough to change the household's actual division of labor, not when it moves only slightly. That is important for interpreting the aggregate Korean trend. Fathers' childcare time has risen, but the observed increase may still be too small to meaningfully reduce mothers' burden.

Moreover, the evidence is also not uniformly pro-natalist. \citet{Lee2022PronatalistLeave} shows that fathers' parental leave in South Korea can reduce intentions for additional children rather than increase them. In her mixed-methods analysis, fathers with leave experience were less likely to report intentions for another child, and the interview evidence suggests that leave made the difficulty of childcare more visible in a setting where broader support for combining work and care remains weak. A recent meta-analysis still finds a positive association between husbands' involvement in parenting and fertility intentions, but the estimated effect is modest rather than large \citep{KimEtAl2024FertilityIntentions}. Overall, this literature suggests a more qualified interpretation. Fathers' involvement may matter, but it is unlikely to matter in a simple mechanical way.

The third strand concerns the well-known quantity-quality tradeoff. \citet{AndersonKohler2013} argue that Korea's low fertility is tied in part to the pressure parents feel to raise successful and competitive children in an environment of intense educational competition. \citet{KimTertiltYum2024} formalize this idea through status externalities in education and show that competition in educational investment can depress fertility. This part of the literature matters here because it helps explain why time allocation inside the household may be so important. When each child is expected to require high levels of care and investment, the way mothers' and fathers' time combine becomes central to the cost of fertility.

Though the micro evidence suggests that greater support from fathers can matter for fertility, aggregate time-use evidence points to a more complicated relationship. Over time, fathers have done more childcare, yet mothers have not done less, and fertility has remained extremely low. We do not read this as evidence that fathers' involvement is irrelevant. Rather, we take it as evidence that the type of paternal involvement matters. Additional father time may reduce mothers' burden when it substitutes for maternal care in routine, solo, and responsibility-bearing tasks. But when father time is complementary to mother time, or when mothers remain the default accountable parent, greater paternal involvement may increase total parental care without substantially lowering the effective cost of childbearing. The model we develop in the next section addresses this problem by putting the substitutability of parental time at the center of the analysis. If parental time is highly substitutable, greater paternal involvement could reduce the effective time cost of another child. If parental time is complementary, or only weakly substitutable, then more father time may still leave mothers with most of the burden. In that case, fertility can remain low even when fathers are more involved than before, especially when parents also face strong pressure to invest heavily in each child.

\section{Substitutes or complements?}

We consider an OLG closed economy where agents live for two periods: childhood and adulthood. There is an initial endowment of human capital per person ($k_0$) and an adult population size ($N_0$) living on a fixed quantity of land ($L$). Households face a fundamental trade-off between the quantity and quality of offspring, as raising more children reduces the time and resources available for each child. The structure of the model follows very closely \cite{delacroix_gosseries_2012}, abstracting from the environmental component and instead bringing intra-household time allocation to the center of the stage (see also \citealp{delacroix2002theory}). Adults allot their time between work and child-rearing, with parental time further disaggregated between fathers ($\theta_d$) and mothers ($\theta_m$). We abstract from leisure and domestic labor to focus on the parental care technology, implicitly assuming equal bargaining power within the household.\footnote{This assumption does not undermine the relevance of the literature on asymmetric household bargaining power (e.g. \citealp{LundbergPollak1993}; \citeyear{LundbergPollak1996}; for OLG models in which mothers bear the whole or most of the cost of rearing children, see \citealp{Agenor2017computable}; \citealp{AgenorAg2023Infrastructure}). 
Rather, our aim is to maintain the most parsimonious structure possible 
in order to isolate the mechanism of interest. Explicitly incorporating 
asymmetric bargaining would reinforce, rather than overturn, the results 
obtained here.}

As the intertemporal optimization problem is specified at the household level, we rule out the possibility of divorce and couples are assumed to die together.\footnote{Marriage rates and the timing of marriage are also important for understanding fertility decline, particularly in South Korea where childbirth remains overwhelmingly concentrated within marriage. Recent work shows that Korea's fertility decline has coincided not only with delayed marriage but also with a broader retreat from marriage itself (see \citealp{LeeParkChin2025}, \citealp{RaymoPark2020}; for broader context, \citealp{RaymoEtAl2015}). We do not claim that the channel studied here is the dominant one, but rather that it constitutes a meaningful mechanism linking intra-household time allocation to fertility decisions. Extending the framework to incorporate marriage formation and divorce is left for future research.} Labor income depends on both hours worked and individual human capital. Investment in children's education and the quality of the parental workforce determine the accumulation of human capital across generations, and physical space and parental effort impose additional constraints on fertility. The government intervenes through a fertility subsidy financed by an income tax, but households retain discretion over how to allocate resources to their children.

At each period $t$, there is a new adult generation of size $N$ that derives utility from consumption ($c$), the number of children ($n$), and their quality, which is captured by their future human capital ($k$). That is:
\[
U=\ln c_{t}+\ln n_{t}k_{t+1}%
\]
where, for simplicity, we assume consumption and being parents provide the same utility, though there is a trade-off between the number and quality of the offspring.

The household budget constraint states that their consumption plus education spending per child ($e$) must be equal to their net income:
\begin{equation}
c_{t}+n_{t}e_{t}=\left(  1-\tau\right)  y_{t}+B(n_{t})
\label{Budget_constraint}%
\end{equation}
where $B(\cdot)$ is an increasing function in $n$ that stands for the fertility subsidy, and $\tau>0$ is the tax rate that funds
the latter. The government has no direct control over how households will spend the subsidy.

We consider two alternative treatments of the government budget. In both of them, the budget is balanced at all times so that fertility subsidies are fully funded by an income tax. However, in the first case, the household fully internalizes Eq. (\ref{Gov_budget}), while in the second case, it does not. We begin with the treatment in which this constraint is fully incorporated into the household optimization problem. This assumption simplifies the analysis, allowing us to show in a more transparent way the mechanism driving our main results on parental time being substitutes or complements. Thus, we have:
\begin{equation}
B(n_{t})=\tau y_{t} \label{Gov_budget}%
\end{equation}
so that the fertility subsidy is fully funded by income taxes. In the second treatment, which we discuss in Section 4, the government adopts an ex post approach: it chooses the tax rate required to balance the budget after households have made their allocation decisions. This setup allows us to isolate more clearly the potentially distorting effects of fertility subsidies on the quantity and quality of children.

Households also face a time constraint. They can allocate their time either to working or non-working activities. In the context of our model, given that we abstract from leisure, adults either work or parent:
\begin{equation}
h_{t}+\theta_{t}=1 \label{Time_constraint}%
\end{equation}
where $h$ is the number of hours worked and $\theta$ stands for the number of
hours dad and mom put together to raise the offspring, so that:%
\begin{equation}
\theta_{t}=\theta_{d,t}+\theta_{m,t} \label{Time_parents}%
\end{equation}
An important distinction of our model relative to previous studies on parental time allocation (e.g. \citealp{Agenor2021GenderGaps}; \citealp{HeintzFolbre2022}; \citealp{AgenorAg2023Infrastructure}) is that we do not impose gender-based specialization in child-rearing. In much of the literature, mothers are assumed to bear the entirety or the majority of the time cost associated with raising children, effectively specializing in this activity within the household (\citealp{Agenor2017computable}). By contrast, we allow both parents to contribute to child-rearing without exogenous gender-based constraints. Such a simplification is not intended to deny the existence or importance of gender differences in parental time use, but rather to highlight that the main mechanism driving our results operates independently of any assumed gender discrimination.

We assume the adult population is self-employed, so output equals their income immediately. The productivity of each hour of work is determined by the worker's quality:
\begin{equation}
y_{t}=h_{t}k_{t} \label{Production_function}%
\end{equation}
From which it follows that there are two possible channels through which income can increase. Either at the extensive margin through longer working hours or by increasing the quality of labor. Still, the latter depends on decisions made by the previous generation.

Empirical evidence suggests that higher-educated parents have a stronger preference for children's skills or perceive greater returns to investing in those skills (\citealp{Cunha2020}). Thus, we assume future human capital is driven by education spending and the present quality of the workforce:
\begin{equation}
k_{t+1}=A e_{t}^{\alpha}k_{t}^{\beta}\label{k_dyn}%
\end{equation}
where $A>0$ is a technological constant, and $0<\alpha+\beta<1$ so that we have decreasing returns. While the assumption rules out endogenous growth in the current setup, the framework is sufficiently general to be extended to incorporate it.

Producing offspring requires time-effort, and space:
\[
n_{t}=\frac{\Theta_{t}}{k_{t}}\left(  \frac{L}{\phi N_{t}}\right)  ^{\gamma}%
\]
where $\Theta$ corresponds to the \textit{effective} time units necessary to raise a child, $L/N$ stands for land or physical space per adult, $\phi>0$ is a scaling parameter, and $0<\gamma <1$ captures its decreasing returns. The term $\Theta/k$ captures the increasing opportunity cost of having children for young adult cohorts with higher human capital. To simplify the algebraic steps, let us normalize $L$, so that the expression above can be rewritten as:
\begin{equation}
k_{t}n_{t}=\frac{\Theta_{t}}{{\phi}^{\gamma} N_{t}^{\gamma}}\label{Offspring_cost}%
\end{equation}

The variable $\Theta$ consists of a key innovation of our model because it allows for the distinction between actual parental hours and effective time dedicated to raising a child. There is an important distinction between inputs into child development that require direct parental involvement, including activities such as reading, sports, or cultural outings, and external childcare services in which children are cared for independently of their parents (see \citealp{CaucuttLochnerMullinsPark2026}). We focus on the former, acknowledging that while parents may allocate a certain number of hours to childcare, these hours do not translate one-to-one into effective input. Effectiveness depends on factors such as parental coordination, their skills and effort. By explicitly modeling $\Theta$ as the effective time units necessary to raise a child, we are able to capture how variations in parental contributions and the complementarity or substitutability of their time affect the quantity-quality trade-off in fertility decisions. Thus, suppose the following CES function:
\begin{equation}
\Theta_{t}=\left(  \theta_{d,t}^{\rho}+\theta_{m,t}^{\rho}\right)  ^{\frac
{1}{\rho}}\label{Effective_time}%
\end{equation}
where $\rho$ is the substitution parameter. If $\rho=0$, the elasticity of substitution between the time dad and mom dedicate to their child is equal to one. When $\rho=1$, they are perfect substitutes. As $\rho\rightarrow-\infty$, parents' time becomes complementary to each other.

Note that $\rho$ should not be interpreted purely as a preference parameter over the combination of parental time inputs in childcare. It also embodies social and institutional norms, and the set of skills that mothers and fathers bring to child-rearing tasks. High substitutability reflects a situation in which both parents are capable of performing the same household tasks interchangeably. As the elasticity of substitution falls, parents become increasingly specialized. This is not reducible to preferences, as it depends on the intra-household cultural transmission of skills across generations. To give a practical example, if a father can cook but cannot drive, and a mother can drive but cannot cook, the two become highly complementary. If both parents can cook and drive, they are substitutes.

Substituting Eq. (\ref{Effective_time}) into (\ref{Offspring_cost}) and rearranging, we can write the time allocated by the mother as a function of the number of children and the time the father spends with them:
\begin{equation}
\theta_{m,t}=\left(  k_{t}^{\rho}n_{t}^{\rho}\phi^{\rho\gamma} N_{t}^{\rho\gamma}-\theta
_{d,t}^{\rho}\right)  ^{\frac{1}{\rho}}\label{Time_mother}%
\end{equation}
We choose to write $\theta_d$ on the right-hand side of the expression to explicitly capture the idea that mothers adjust their time allocation in response to the hours fathers dedicate to childcare. This is how societal norms enter the problem and are reflected in the technology. Notice that for $\rho>0$, higher fertility requires mothers to dedicate more time to non-working activities, but this effect is balanced by the time fathers allocate to raising their offspring. Thus, there is a substitution between parental labor in child-rearing. However, for $\rho<0$, an increase in $\theta_d$ necessarily increases $\theta_m$. Such a difference will serve as the transmission channel driving our main results.

By substituting Eqs. (\ref{Time_mother}) into (\ref{Time_parents}), and the resulting expression into (\ref{Time_constraint}), we can rewrite the time constraint in terms of the number of hours worked as:
\begin{equation}
h_{t}=1-\theta_{d,t}-\left(  k_{t}^{\rho}n_{t}^{\rho}\phi^{\rho\gamma} N_{t}^{\rho\gamma}%
-\theta_{d,t}^{\rho}\right)  ^{\frac{1}{\rho}}\label{Hours_worked}%
\end{equation}
Having children reduces the time parents can devote to work, but the effect is larger the lower the substitutability between mothers and fathers. If we substitute Eq. (\ref{Hours_worked}) into our production function in (\ref{Production_function}), the choice of $h$ is translated into production:
\begin{equation}
y_{t}=\left[  1-\theta_{d,t}-\left(  k_{t}^{\rho}n_{t}^{\rho}\phi^{\rho\gamma} N_{t}^{\rho
\gamma}-\theta_{d,t}^{\rho}\right)  ^{\frac{1}{\rho}}\right]  k_{t}%
\label{Production_function_parents}%
\end{equation}

Making use of Eq. (\ref{Production_function_parents}) and the government budget constraint in (\ref{Gov_budget}), we can rewrite the household budget constraint (\ref{Budget_constraint}) in terms of consumption as:%
\begin{equation}
c_{t}=\left[  1-\theta_{d,t}-\left(  k_{t}^{\rho}n_{t}^{\rho}\phi^{\rho\gamma} N_{t}^{\rho
\gamma}-\theta_{d,t}^{\rho}\right)  ^{\frac{1}{\rho}}\right]  k_{t}-n_{t}%
e_{t}\label{Consumption}%
\end{equation}

Thus, the household intertemporal problem consists in:
\begin{align}
&  \underset{c_{t},n_{t},e_{t}}{\max}\ln c_{t}+\ln n_{t}k_{t+1}%
\label{Optimisation_1}\\
&  \text{s.t.}\nonumber\\
c_{t} &  =\left[  1-\theta_{d,t}-\left(  k_{t}^{\rho}n_{t}^{\rho}\phi^{\rho\gamma} N_{t}%
^{\rho\gamma}-\theta_{d,t}^{\rho}\right)  ^{\frac{1}{\rho}}\right]
k_{t}-n_{t}e_{t}\nonumber
\end{align}
where agents choose their consumption, the quantity and quality of their offspring. Substituting the constraint into the objective function in (\ref{Optimisation_1}), and making use of Eq. (\ref{k_dyn}), we can rewrite the lifetime choice as:%
\begin{align}
&  \underset{n_{t},e_{t}}{\max}\ln\left\{  \left[  1-\theta_{d,t}-\left(
k_{t}^{\rho}n_{t}^{\rho}\phi^\rho N_{t}^{\rho\gamma}-\theta_{d,t}^{\rho}\right)
^{\frac{1}{\rho}}\right]  k_{t}-n_{t}e_{t}\right\}  \label{Optimisation_2}\\
&  \text{
\ \ \ \ \ \ \ \ \ \ \ \ \ \ \ \ \ \ \ \ \ \ \ \ \ \ \ \ \ \ \ \ \ \ \ \ \ \ \ \ \ \ \ }%
+\ln n_{t}+\ln A+\alpha\ln e_{t}+\beta\ln k_{t}\nonumber
\end{align}
so that we end up with an unconstrained optimization problem in which the choice variables are strictly related to the quantity--quality trade-off.

Along the dynamic equilibrium path, the First Order Conditions (FOC) are:
\begin{align}
\frac{-\left[  1-\left(  \frac{\theta_{d,t}}{k_{t}n_{t}\phi N_{t}^{\gamma}}\right)
^{\rho}\right]  ^{\frac{1-\rho}{\rho}}k_{t}-e_{t}}{\left[  1-\theta
_{d,t}-\left(  k_{t}^{\rho}n_{t}^{\rho}\phi^\rho N_{t}^{\rho\gamma}-\theta_{d,t}^{\rho
}\right)  ^{\frac{1}{\rho}}\right]  k_{t}-n_{t}e_{t}}+\frac{1}{n_{t}} &
=0\nonumber\\
& \label{FOC_general}\\
-\frac{n_{t}}{\left[  1-\theta_{d,t}-\left(  k_{t}^{\rho}n_{t}^{\rho}%
\phi^\rho N_{t}^{\rho\gamma}-\theta_{d,t}^{\rho}\right)  ^{\frac{1}{\rho}}\right]
k_{t}-n_{t}e_{t}}+\frac{\alpha}{e_{t}} &  =0\nonumber
\end{align}

\begin{figure}[htbp]
\centering
\begin{tikzpicture}[
    x=1cm,y=1cm,
    >=stealth,
    box/.style={
        draw=black!70,
        thick,
        rectangle,
        rounded corners=2pt,
        align=center,
        inner sep=6pt,
        minimum height=1.2cm
    },
    bluebox/.style={box, fill=blue!18},
    greenbox/.style={box, fill=green!25},
    redbox/.style={box, fill=red!18},
    arrow/.style={->, thick, draw=black!75}
]

\node[bluebox, text width=5.8cm] (hh) at (0,0)
{Household intertemporal maximization\\[3pt]
{\footnotesize $\max \ \ln c_t + \ln(n_t k_{t+1})$}};

\node[greenbox, text width=6.3cm] (theta) at (0,-2.6)
{Parental-care technology\\[3pt]
{\footnotesize $\Theta_t=\left(\theta_{d,t}^{\rho}+\theta_{m,t}^{\rho}\right)^{1/\rho}$}};

\node[greenbox, text width=5.4cm] (subs) at (-5.2,-5.8)
{Perfect substitutes\\[3pt]
{\footnotesize $\rho=1$}\\
{\footnotesize father time offsets mother time}\\[3pt]
{\footnotesize higher $n_t$, lower $e_t$}\\
{\footnotesize higher $\bar N$, same $\bar k$}};

\node[greenbox, text width=5.4cm] (comp) at (5.2,-5.8)
{Perfect complements\\[3pt]
{\footnotesize $\rho\rightarrow -\infty$}\\
{\footnotesize father and mother time move together}\\[3pt]
{\footnotesize lower $n_t$, higher $e_t$}\\
{\footnotesize lower $\bar N$, same $\bar k$}};

\node[redbox, text width=7.2cm] (ss) at (0,-9.2)
{Steady-state comparison\\[3pt]
{\footnotesize $\bar{k}_S=\bar{k}_C$}\\
{\footnotesize $\bar{N}_C=\Omega \bar{N}_S,\qquad \Omega=2^{-1/\gamma}\in(0,1)$}};

\draw[arrow] (hh.south) -- (theta.north);
\draw[arrow] (theta.south west) -- (subs.north);
\draw[arrow] (theta.south east) -- (comp.north);
\draw[arrow] (subs.south) -- ([xshift=-1cm]ss.north);
\draw[arrow] (comp.south) -- ([xshift=1cm]ss.north);

\end{tikzpicture}
\caption{Main elements of the model. The parental-care technology determines whether maternal and paternal time are perfect substitutes or perfect complements, which in turn shapes fertility, education, and the steady-state comparison.}
\label{fig:model_transmission}
\end{figure}
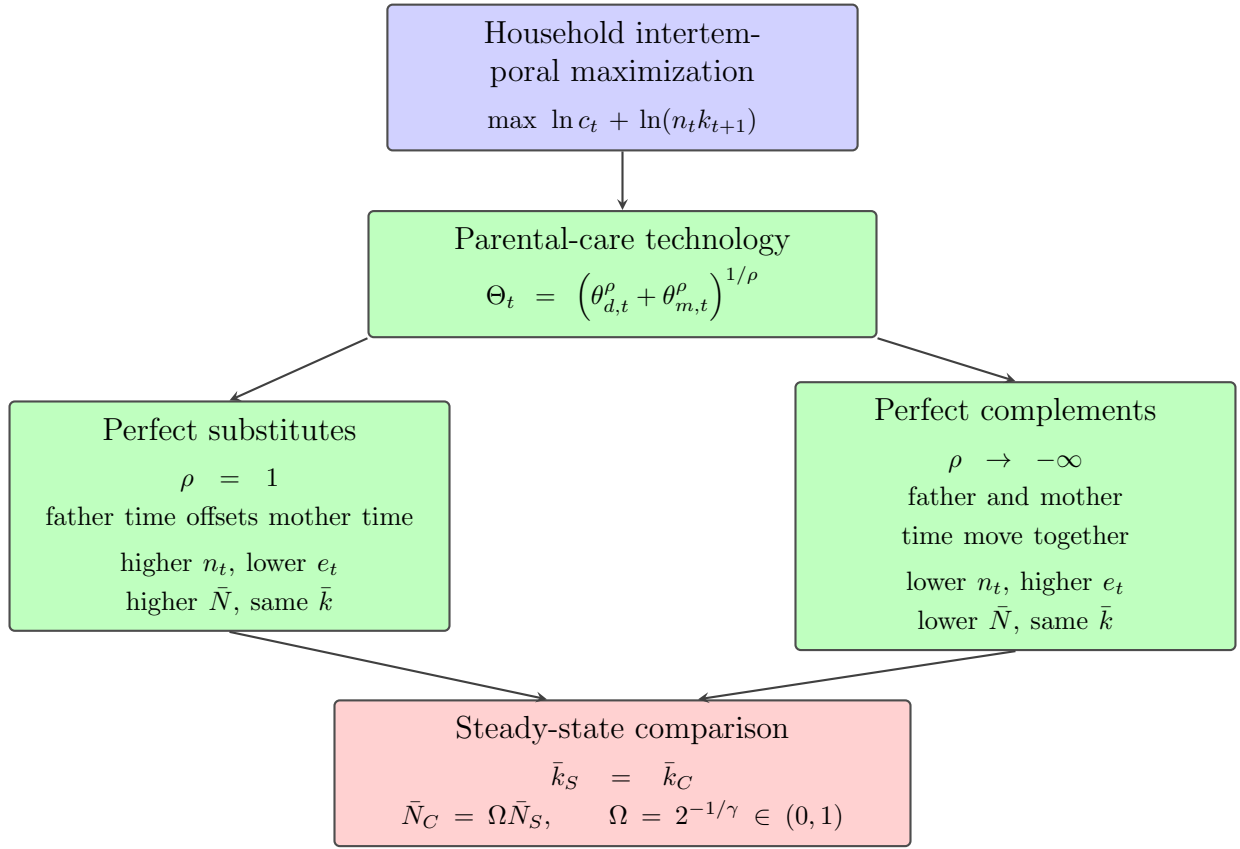

The 2-dimensional nonlinear dynamic system is defined by Eq. (\ref{k_dyn}) and the evolution of the adult population, both of which depend on the households' optimal choices.
\begin{align}
k_{t+1} &  =A e_{t}^{\ast \alpha}k_{t}^{\beta
}\nonumber\\
& \label{DynamSys_general}\\
N_{t+1} &  =n_{t}^{\ast}N_{t}\nonumber
\end{align}
where $e^*$ and $n^*$ describe the optimal allocations satisfying (\ref{FOC_general}). Given the strong nonlinearities in the first-order conditions, our strategy will be to analyze the model both analytically and through numerical simulations by focusing on two extreme cases: first, when parental time is perfectly substitutable, and second, when it is fully complementary. For each adult generation, the initial human capital and population size are treated as exogenous. Conditional on these values, we can determine households' optimal choices for fertility and educational investment. As each generation passes, these decisions feed back into the aggregate human capital and population size, shaping the dynamic evolution of the economy over time. We will discuss both micro and macro dimensions for each case. Fig. \ref{fig:model_transmission} shows the main elements of the model.

\bigskip

\subsection{Perfect substitution ($\rho=1$)}

When there is perfect substitution between the time of dad and mom, the FOC become:
\begin{align}
\frac{-k_{t}^{2}\phi N_{t}^{\gamma}-e_{t}}{\left(  1-\theta_{d,t}-n_{t}k_{t}%
\phi N_{t}^{\gamma}+\theta_{d,t}\right)  k_{t}-n_{t}e_{t}}+\frac{1}{n_{t}} &
=0\nonumber\\
& \label{FOC_subs}\\
\frac{-n_{t}}{\left(  1-\theta_{d,t}-n_{t}k_{t}\phi N_{t}^{\gamma}+\theta
_{d,t}\right)  k_{t}-n_{t}e_{t}}+\frac{\alpha}{e_{t}} &  =0\nonumber
\end{align}

Provided that $k_{t}>0$, the optimal fertility and education choices are:%
\begin{align*}
n_{t}^{\ast} &  =\frac{1-\alpha}{2k_{t}\phi N_{t}^{\gamma}}\\
e_{t}^{\ast} &  =\left(  \frac{\alpha}{1-\alpha}\right)  k_{t}^{2}%
\phi N_{t}^{\gamma}%
\end{align*}
Both expressions are intuitive and consistent with standard economic reasoning. When a household faces a macroeconomic environment characterized by a high aggregate human capital stock and a relatively large population with respect to the fixed quantity of land, both the opportunity and space costs of raising children increase. As a result, the optimal fertility rate declines. At the same time, these higher costs create a stronger incentive for parents to invest in the quality of each child, leading to an increase in optimal education investment.

Substituting the optimal choices into (\ref{DynamSys_general}), the dynamic system is given by:
\begin{align}
k_{t+1} &  =A\left[  \left(  \frac{\alpha}{1-\alpha}\right)  k_{t}^{2}%
\phi N_{t}^{\gamma}\right]  ^{\alpha}k_{t}^{\beta}\nonumber\\
& \label{Dynam_Sys_subs}\\
N_{t+1} &  =\left(  \frac{1-\alpha}{2\phi k_{t}}\right)
N_{t}^{1-\gamma}\nonumber
\end{align}
In steady-state, $k_{t+1}=k_{t}=\bar{k}$ and $N_{t+1}=N_{t}=\bar{N}\,$. This
results in the following equilibrium conditions:
\begin{align}
\bar{k} &  =A\left[  \left(  \frac{\alpha}{1-\alpha}\right)  \bar{k}^{2}%
\phi \bar{N}^{\gamma}\right]  ^{\alpha}\bar{k}^{\beta}\nonumber\\
& \label{Equilibrium_subs}\\
\bar{N} &  =\left(  \frac{1-\alpha}{2\phi\bar{k}}\right)  \bar
{N}^{1-\gamma}\nonumber
\end{align}

Thus, we can state and prove the following Proposition regarding the existence of a unique non-trivial equilibrium solution.

\begin{proposition} \label{prop 1}
When the time allocated by father and mother to parenting are perfect substitutes, the dynamic system (\ref{Dynam_Sys_subs}) has a unique economically meaningful equilibrium point $P_S=(\bar{k_S},\bar{N_S})$ defined and given by:
\begin{equation*}
\bar{k_S}=\left(  A^{\frac{1}{\alpha}}\frac{\alpha}{2}\right)  ^{\frac{\alpha
}{1-\beta-\alpha}}
\end{equation*}
\begin{equation*}
\bar{N_S}=A^{\frac{-1}{\left(  1-\beta-\alpha\right)  \gamma}}\underset
{\text{{\scriptsize Fertility}}}{\underbrace{\left(  \frac{1-\alpha}{2\phi
}\right)  }}^{\frac{1-\beta-2\alpha}{\left(  1-\beta-\alpha\right)  \gamma}%
}\underset{\text{{\scriptsize Education}}}{\underbrace{\left[  \left(
\frac{\alpha}{1-\alpha}\right)  \phi\right]  }}^{\frac{-\alpha}{\left(
1-\beta-\alpha\right)  \gamma}}%
\end{equation*}

\bigskip

\begin{proof}
See Appendix A.1.
\end{proof}

\end{proposition}

\medskip

\noindent Notice that it is possible to disaggregate the equilibrium population size into its fertility and education components. Both are linked to the returns from education spending on human capital for the future generation ($\alpha$). Regarding the local stability of the steady-state solution, we are now ready to prove the following Proposition.

\medskip

\begin{proposition} \label{prop 2}
The unique non-trivial equilibrium $P_S=(\bar{k_S},\bar{N_S})$ of the dynamic system (\ref{Dynam_Sys_subs}) is locally stable provided that:
\begin{align}
1-\frac{\alpha\frac{\partial e^{\ast\alpha-1}}{\partial\bar{N}}\bar{k}^{\beta
}\frac{\partial n^{\ast}}{\partial\bar{k}}\bar{N}}{\alpha\frac{\partial
e^{\ast\alpha-1}}{\partial\bar{k}}\bar{k}^{\beta}+\frac{\beta}{A}}-\frac
{1}{A\alpha\frac{\partial e^{\ast\alpha-1}}{\partial\bar{k}}\bar{k}^{\beta
}+\beta} &  <-\frac{\partial n^{\ast}}{\partial\bar{N}}\bar{N}<1-\frac
{A\alpha\frac{\partial e^{\ast\alpha-1}}{\partial\bar{N}}\bar{k}^{\beta}%
\frac{\partial n^{\ast}}{\partial\bar{k}}\bar{N}}{1+A\alpha\frac{\partial
e^{\ast\alpha-1}}{\partial\bar{k}}\bar{k}^{\beta}+\beta} \nonumber \\ \label{stab_subs} \\
A\alpha\frac{\partial e^{\ast\alpha-1}}{\partial\bar{k}}\bar{k}^{\beta}+\beta
&  >1 \nonumber
\end{align}
where
\begin{align*}
    \frac{\partial n^{*}}{\partial \bar{k}} &= -\frac{1-\alpha}{2 \bar{k}^{2}
\bar{N}^{\gamma}}, \qquad \qquad \frac{\partial n^{*}}{\partial \bar{N}} = -\frac{\gamma(1-\alpha)}{2 \bar{k}
\bar{N}^{\gamma+1}}, \\
\frac{\partial e^{*}}{\partial \bar{k}} &= \frac{2\alpha}{1-\alpha} \, \bar{k}
\bar{N}^{\gamma}, \qquad \quad \frac{\partial e^{*}}{\partial \bar{N}} = \frac{\alpha\gamma}{1-\alpha} \,
\bar{k}^{2} \bar{N}^{\gamma-1}
\end{align*}

\bigskip

\begin{proof}
See Appendix A.2.
\end{proof}

\end{proposition}

\medskip

The Proposition above identifies two conditions that must hold jointly for the equilibrium to be locally stable, each with a clear economic interpretation. The first concerns the responsiveness of fertility decisions to population size, a relationship mediated by land availability. If households are unresponsive to changes in available space, they do not sufficiently adjust their fertility decisions as land becomes scarce, and the system loses its self-correcting properties. But the opposite extreme is equally destabilizing. Households that are too sensitive to land scarcity may sharply curtail fertility at the first sign of congestion, pushing the economy toward demographic collapse. Therefore, stability requires that this elasticity falls within a corridor bounded by a floor and a ceiling. Too little responsiveness and the population dynamic diverges from above; too much and it collapses from below.

The second condition concerns the intergenerational transmission of human capital. For the equilibrium to be stable, optimal education investment by the current generation must be sufficiently responsive to the human capital stock inherited from the previous one. If parental education has no meaningful effect on how much the current generation invests in its children's schooling, the intergenerational bridge sustaining the education dynamic is broken, and the system again becomes unstable. The two conditions reflect a deeper feature of the model. Fertility and education decisions are made by young adults who respond to state variables, namely population size and the aggregate human capital stock, that were determined by the choices of the generation before them. Stability requires that both margins of adjustment, fertility and education investment, be calibrated so that this inter-generational feedback loop neither explodes nor collapses.

\subsection{Perfect complements ($\rho=-\infty$)}

Perfect substitutability and perfect complementarity represent the two boundaries of the substitution spectrum. Having analyzed the first, we turn to the second. When the time of dad and mom are complements, the FOCs become:
\begin{align}
\frac{-2k_{t}^{2}\phi N_{t}^{\gamma}-e_{t}}{\left(  1-2n_{t}k_{t}\phi N_{t}^{\gamma
}\right)  k_{t}-n_{t}e_{t}}+\frac{1}{n_{t}} &  =0\nonumber\\
& \label{FOC_comp}\\
\frac{-n_{t}}{\left(  1-2n_{t}k_{t}\phi N_{t}^{\gamma}\right)  k_{t}-n_{t}e_{t}%
}+\frac{\alpha}{e_{t}} &  =0\nonumber
\end{align}

If $k_{t}>0$, the optimal fertility and education choices are:%
\begin{align*}
n_{t}^{\ast} &  =\frac{1-\alpha}{4k_{t}\phi N_{t}^{\gamma}}\\
e_{t}^{\ast} &  =2\left(  \frac{\alpha}{1-\alpha}\right)  k_{t}^{2}%
\phi N_{t}^{\gamma}%
\end{align*}
The optimal allocations under perfect complementarity differ strikingly from those obtained under perfect substitutability. Fertility is half as large, and education investment is twice as high. The intuition follows directly from the structure of the effective parental time function ($\Theta$). Under perfect complementarity, fathers and mothers must contribute jointly to childcare. An increase in paternal involvement necessarily requires a corresponding increase in maternal time at home. This crowds out time that would otherwise be devoted to market work, putting downward pressure on household income. Faced with this tighter constraint, households optimally respond by reducing the number of children. The resulting loss in utility from lower fertility is then offset by investing more heavily in the quality of each child.

Substituting once more the optimal allocations into (\ref{DynamSys_general}), the dynamic system is given by:
\begin{align}
k_{t+1} &  =A\left[  2\left(  \frac{\alpha}{1-\alpha}\right)  k_{t}^{2}%
\phi N_{t}^{\gamma}\right]  ^{\alpha}k_{t}^{\beta}\nonumber\\
& \label{Dynam_Sys_comp}\\
N_{t+1} &  =\left(  \frac{1-\alpha}{4\phi k_{t}}\right)
N_{t}^{1-\gamma}\nonumber
\end{align}
In steady-state, $k_{t+1}=k_{t}=\bar{k}$ and $N_{t+1}=N_{t}=\bar
{N}\,$. This results in the following equilibrium conditions:%
\begin{align}
\bar{k} &  =A\left[  2\left(  \frac{\alpha}{1-\alpha}\right)  \bar{k}^{2}%
\phi \bar{N}^{\gamma}\right]  ^{\alpha}\bar{k}^{\beta}\nonumber\\
& \label{Equilibrium_comp}\\
\bar{N} &  =\left(  \frac{1-\alpha}{4\phi \bar{k}}\right)  \bar
{N}^{1+\gamma}\nonumber
\end{align}

We proceed by stating and proving the following Proposition regarding the existence of a unique non-trivial equilibrium solution.
\begin{proposition} \label{prop 3}
When the time allocated by father and mother to parenting are perfect complements, the dynamic system (\ref{Dynam_Sys_comp}) has a unique economically meaningful equilibrium point $P_{C}=(\bar{k_C},\bar{N_C})$ defined and given by:

\begin{equation*}
\bar{k_C}=\bar{k_S} \qquad \qquad  \bar{N_C}=\Omega\bar{N_S}
\end{equation*}
where
\begin{align*}
\Omega  & =\frac{1}{2^{1/\gamma}} \in (0,1)
\end{align*}

\bigskip

\begin{proof}
See Appendix A.3.
\end{proof}
\end{proposition}
Perfect complementarity in parenting time produces, at the aggregate level, the same level of human capital as before, but a lower $\bar{N}$. The decline in population is intuitive, since fertility is lower in this case. The human capital result, however, is less obvious. It follows from our assumption of decreasing returns to education spending. Additional schooling expenditures raise human capital, but at a diminishing rate.

\begin{proposition} \label{prop 4}
The unique non-trivial equilibrium solution $P_C=(\bar{k_C},\bar{N_C})$ of the dynamic system (\ref{Dynam_Sys_comp}) is locally stable provided that (\ref{stab_subs}) is satisfied, and we further have:
\begin{align*}
    \frac{\partial n^*}{\partial \bar{k}}
&= -\frac{1-\alpha}{4 \bar{k}^2 \bar{N}^{\gamma}}, \qquad \qquad \frac{\partial n^*}{\partial \bar{N}}
= -\frac{\gamma (1-\alpha)}{4 \bar{k} \bar{N}^{\gamma+1}}, \\
\frac{\partial e^*}{\partial \bar{k}}
&= \frac{4\alpha}{1-\alpha} \bar{k} \bar{N}^{\gamma}, \qquad \quad \frac{\partial e^*}{\partial \bar{N}}
= \frac{2\alpha\gamma}{1-\alpha} \bar{k}^2 \bar{N}^{\gamma-1}
\end{align*}

\bigskip

\begin{proof}
See Appendix A.4.
\end{proof}

\end{proposition}

The economic interpretation of the stability conditions mirrors that of the perfect substitutes case in broad structure but differs quantitatively. As before, stability requires that the elasticity of optimal fertility with respect to population size falls within a bounded corridor. We also need the education investment to respond sufficiently to the aggregate human capital stock inherited from the previous generation. The novelty lies in how tight those requirements are. Under perfect complementarity, the admissible corridor for the fertility elasticity is half as wide as in the substitutes case. Both the floor and the ceiling are halved. At the same time, the minimum degree of responsiveness required from education investment is twice as large. Complementarity, by binding maternal and paternal time together, amplifies the consequences of any misalignment in household decisions. When parents cannot substitute for one another, the system has less room to absorb deviations from the equilibrium path, and stability demands more precisely calibrated responses on both margins.

\subsection{A numerical comparison}

To provide a more concrete view of the model's properties, we complement the analytical results with a set of numerical experiments. Parameter values are loosely informed by \citet{delacroix_gosseries_2012} and chosen to yield equilibrium values broadly consistent with South Korea's demographic reality. These two sources serve only as reference points to discipline the calibration strategy; the parameter values used in the simulations are not structural estimates and do not necessarily coincide with either source. Throughout the analysis, $k$ can be interpreted as years of schooling, and $N$ as the population size in millions.

\begin{figure}[tbp]
    \centering
    \includegraphics[width=0.5\linewidth]{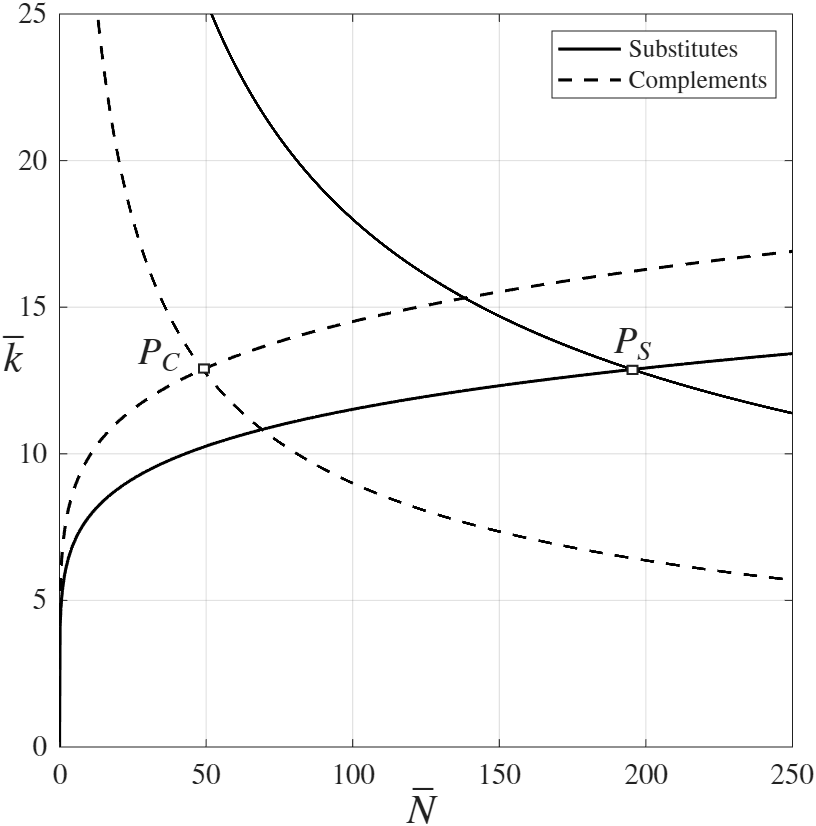}
    \caption{Comparing equilibria when dad and mom's time allocated to parenting are perfect substitutes vs complements. Parameters $A=3.75$, $\alpha=0.1$, $\beta=0.5$, $\gamma=0.5$, $\phi= 0.0025$.}
    \label{Equilibria}
\end{figure}

Fig.~\ref{Equilibria} summarizes the main findings in the $(k, N)$ space. The degree of substitutability between parental time inputs shapes educational investment within a generation but leaves aggregate human capital broadly unchanged at the macro level. This result follows from the presence of decreasing returns to skill accumulation, which absorb differences in per-child investment across time. Population size, however, tells a different story. Perfect substitutability between parental time inputs supports higher fertility, which compounds across generations into a substantially larger equilibrium population. Complementarity, by contrast, depresses fertility and leads to a smaller steady-state population.

\begin{figure}[tbp]
    \centering
    \subfloat[]{\includegraphics[width=3.125in]{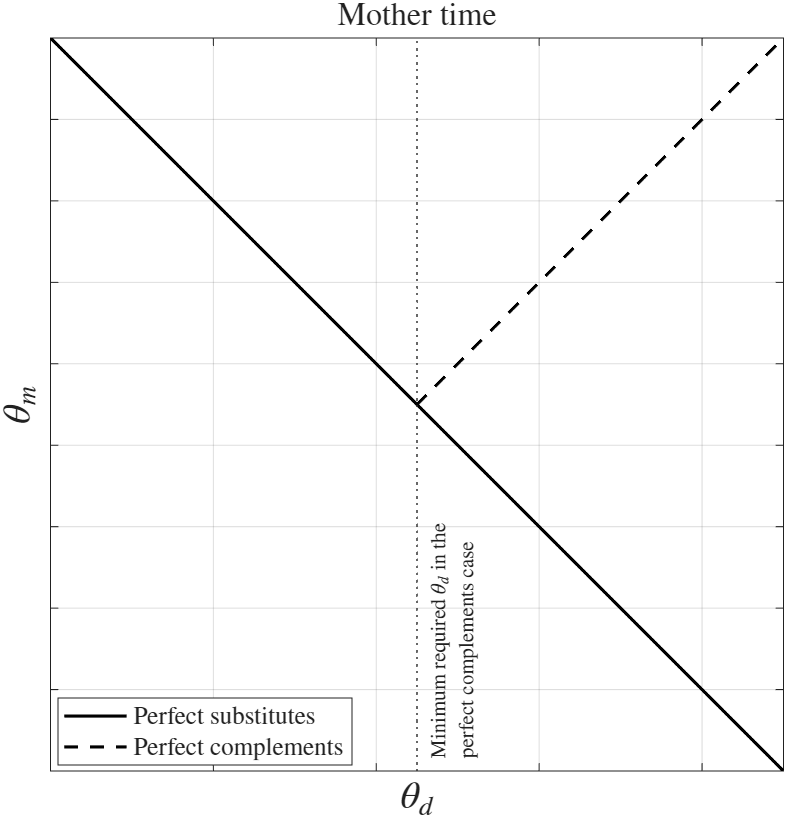}} \quad
    \subfloat[]{\includegraphics[width=3.1in]{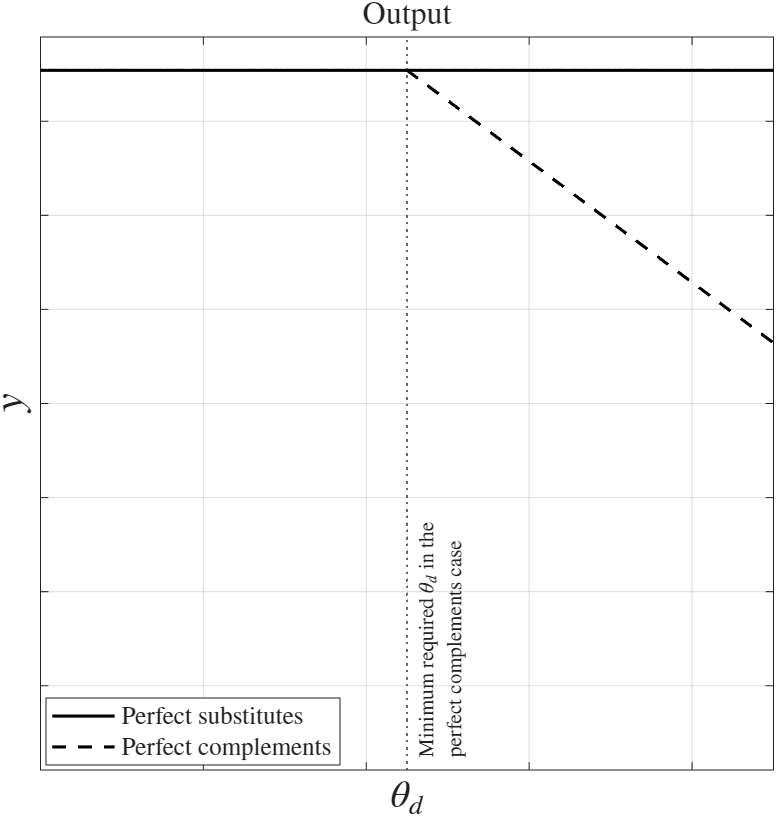}}
    \caption{A comparison of intra-household time allocation and the corresponding output implications.}
    \label{TimeOutput}
\end{figure}

In Fig.~\ref{TimeOutput}, we compare the perfect substitution and perfect complementarity cases in terms of how mothers and fathers allocate their time and the implications for steady-state output. Under substitutability, panel (a) shows that as fathers devote more time to childcare, mothers are able to reduce theirs correspondingly, leaving total parental time unchanged, as indicated by the continuous black line. Panel (b) confirms that this reallocation has no effect on output, which remains stable across all combinations of $\theta_d$ and $\theta_m$. The picture changes markedly under complementarity. Two insights emerge. First, there exists a minimum level of $\theta_d$ below which the model breaks down. Below that, households prefer not to have children, and the economy eventually collapses. Beyond this threshold, panel (a) shows that any increase in fathers' childcare time necessarily requires mothers to increase theirs as well, marked by the dotted black line. This crowds out time from market work, reducing output as shown in panel (b), and exacerbates the quantity-quality trade-off. Faced with tighter time and income constraints, households respond by investing more heavily in each child's education while further reducing fertility.

\begin{figure}[tbp]
  \centering
  \subfloat[]{\includegraphics[width=3.05in]{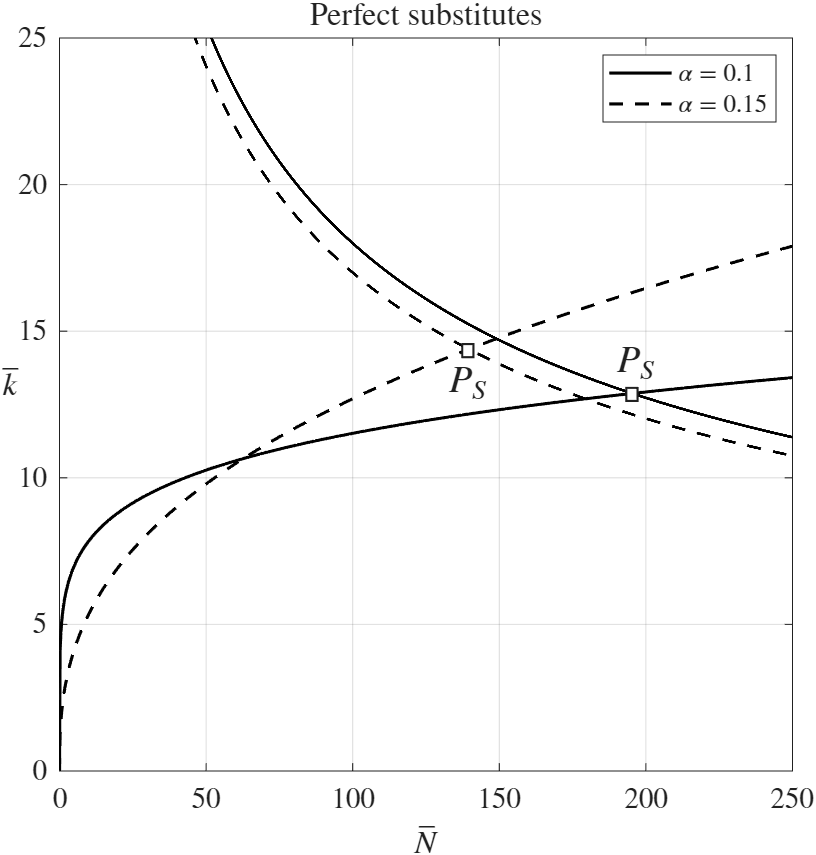}} \hspace{0.75cm}
  \subfloat[]{\includegraphics[width=3.05in]{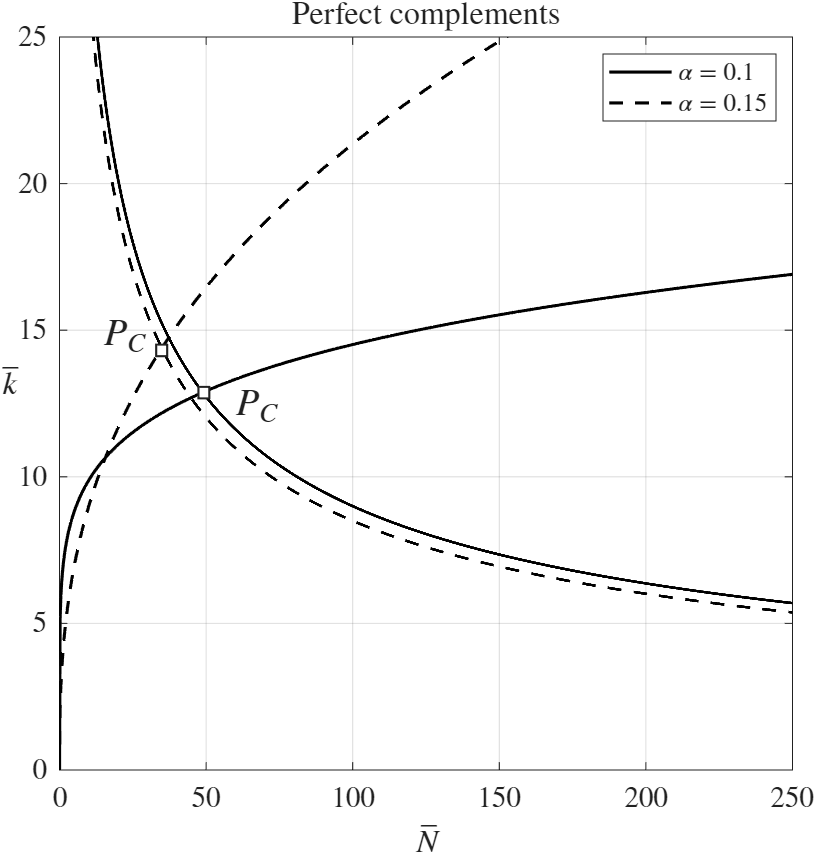}} \\
  \caption{Comparing equilibria for different education returns. Parameters $A=3.75$, $\beta=0.5$, $\gamma=0.5$, $\phi= 0.0025$.}\label{Equilibria_alpha}
\end{figure}

We proceed by reporting the sensitivity of the equilibrium to changes in the return to education ($\alpha$) in Fig.~\ref{Equilibria_alpha}. Under perfect substitutability, panel (a) shows that raising $\alpha$ from 0.1 to 0.15 generates a modest increase in equilibrium schooling of around two years, but is accompanied by a sharp contraction in population size. Under perfect complementarity, panel (b) shows a human capital response of similar magnitude, but the population adjusts by comparatively little. This asymmetry reflects a fundamental difference in the flexibility of the two regimes. When parental time inputs are substitutable, the household can more freely reallocate time across margins, allowing the system to respond more strongly to changes in structural parameters. Complementarity constrains that reallocation.

\section{Distortionary taxation and fertility subsidies effectiveness}

The results derived so far rest on a modeling choice that, while analytically convenient, limits the scope for fiscal policy to play any independent role. By assuming that households fully internalize the implied tax burden, the model embeds a form of Ricardian equivalence by design. Fertility subsidies are offset by anticipated taxes, leaving allocation decisions unchanged. We now look at the more realistic case where the representative household does not take the government's balanced budget into account. Households now respond to the subsidy as a transfer without fully considering its cost, creating scope for fertility subsidies to distort the quantity and quality margins of childbearing. Whether those distortions work in the intended direction, raising fertility, or against it, is the question this section addresses.

From the household perspective, the tax rate $\tau$ and the subsidy schedule $B(\cdot)$ are taken as exogenous. Thus, the household budget constraint becomes:
\begin{equation}
c_t + n_t e_t = (1-\tau) y_t + B(n_t),
\label{BC_dist}
\end{equation}
where $B'(n)>0$ and $\tau\in(0,1)$. Following our previous strategy, we will focus on the two polar cases for parental time allocation.

\subsection{Perfect substitution ($\rho=1$)}

When parenting time is perfectly substitutable, from Eq. (\ref{Hours_worked}), the offspring production or fertility condition implies working hours are: 
\[
h_t = 1 - k_t n_t \phi N_t^\gamma.
\]
which in turn results in an output:
\begin{align}
y_t &= h_t k_t \nonumber \\ &= k_t - k_t^2 \phi N_t^\gamma n_t.
\label{y_subs_dist}
\end{align}

Define $X_t \equiv k_t^2 \phi N_t^\gamma$ as an auxiliary variable, such that $y_t = k_t - X_t n_t$ and $\partial y_t / \partial n_t = -X_t$. Using Eq. (\ref{BC_dist}), consumption can be written as:
\[
c_t = (1-\tau)(k_t - X_t n_t) + B(n_t) - n_t e_t.
\]

Therefore, the first-order conditions are:
\begin{align}
-\frac{n_t}{c_t} + \frac{\alpha}{e_t} = 0
\nonumber \\
\label{FOC_e_dist} \\
\frac{1}{c_t}\left[-(1-\tau)X_t + B'(n_t) - e_t \right] + \frac{1}{n_t} = 0 \nonumber
\end{align}

Combining both expressions in (\ref{FOC_e_dist}) and making use of the budget constraint (\ref{BC_dist}), the optimal fertility choice and education spending are:
\begin{align}
    n_t^\ast=\frac{(1-\alpha)\big[(1-\tau)k_t + B(n_t^\ast)\big]}
{2(1-\tau)X_t - (1+\alpha)B'(n_t^\ast)} \nonumber \\
\label{n_star_subs_dist} \\
e_t^\ast=\frac{\alpha}{1-\alpha}\big[(1-\tau)X_t - B'(n_t^\ast)\big] \nonumber
\end{align}
The two expressions above reveal the intended consequences of the fertility subsidy. A higher $B(\cdot)$ and $B'$ leads households to increase the number of children. Given the trade-off between the quantity and quality of offspring, this is accompanied by a reduction 
in resources allocated to education. As shown in Section 3, however, a lower $e^*$ does not necessarily imply a smaller human capital stock in the steady state, which at first 
sight makes a case in favor of the subsidy. The critical step is how the government balances its budget.

\subsubsection{Endogenizing the tax rate}

To simplify the algebra and without loss of generality, suppose the subsidy takes a linear form:
\[
B(n_t) = b n_t, \qquad 0 < b < 1,
\]
where $b$ corresponds to the fraction of the cost of each child 
covered by the government.

Along the equilibrium path, the government budget constraint
(\ref{Gov_budget}) implies:
\[
b n_t = \tau_t y_t
\]
so that the government will choose the tax rate that balances the budget after fertility and production decisions have been made. Making use of Eq. (\ref{y_subs_dist}), the endogenous tax rate becomes:
\begin{equation}
\tau_t
=
\frac{b n_t}{k_t - X_t n_t},
\label{tau_subs_endog}
\end{equation}
where recall $X_t = k_t^2 \phi N_t^\gamma$.

Substituting Eq. (\ref{tau_subs_endog}) into the fertility optimal condition (\ref{n_star_subs_dist}) eliminates $\tau_t$ and delivers a quadratic
condition in $n_t$:
\begin{equation}
2X_t^2 n_t^2
+
k_t\big[(\alpha-3)X_t+(1+\alpha)b\big]n_t
+
(1-\alpha)k_t^2
=0.
\label{n_quad_subs}
\end{equation}
The economically meaningful solution satisfies $n_t\ge 0$ and $k_t - X_t n_t>0$.

\subsubsection{Dynamic system with endogenous taxation}

Equation (\ref{n_quad_subs}) yields two real roots. Denoting them by
$n_{1,t}^{\ast}$ and $n_{2,t}^{\ast}$, we have:

\begin{equation*}
n_{i,t}^{\ast}
=
\frac{
k_t\left[(3-\alpha)X_t-(1+\alpha)b \pm \sqrt{\Delta_t^{s}}\right]
}{
4X_t^2
},
\qquad i=1,2
\end{equation*}
where
\begin{equation*}
\Delta_t^{s}
=
\left[(3-\alpha)X_t-(1+\alpha)b\right]^2
-
8(1-\alpha)X_t^2.
\end{equation*}
Hence, there are up to two economically meaningful candidate fertility choices,
provided that $\Delta_t^{s}\geq 0$, $n_{i,t}^{\ast}\geq 0$, and
$k_t-X_t n_{i,t}^{\ast}>0$.

Substituting the endogenous tax rate (\ref{tau_subs_endog}) into the second expression in (\ref{n_star_subs_dist}), the education allocation becomes:
\[
e_{i,t}^\ast
=
\frac{\alpha}{1-\alpha}
\left(
\frac{k_t(X_t-b)-X_t^2 n_{i,t}^\ast}
{k_t - X_t n_{i,t}^\ast}
\right).
\]

Thus, our two-dimensional dynamic system with distorting taxes becomes:
\begin{align}
k_{t+1}
&=
A \left[
\frac{\alpha}{1-\alpha}
\left(
\frac{k_t(X_t-b)-X_t^2 n_{i,t}^\ast}
{k_t - X_t n_{i,t}^\ast}
\right)
\right]^\alpha
k_t^\beta,
\nonumber
\\ \label{Dynam_Sys_subs_b} \\
N_{t+1}
&=
n_{i,t}^\ast N_t.
\nonumber
\end{align}
where $n_{i,t}^\ast(k_t,N_t)$ denotes the feasible root of (\ref{n_quad_subs}). We can state and prove the following Proposition regarding the existence of a non-trivial equilibrium solution. Unlike the benchmark case in Section 3, distortionary taxation implies there might be more than one optimal allocation $(n^*,e^*)$, with steady-state implications for our dynamic system.

\begin{proposition} \label{prop 5}
When paternal and maternal parenting time are perfect substitutes, and the pro-natal subsidy is linear, $B(n_t)=bn_t$, there exists a threshold value $\bar{b}>0$ for which:
\begin{enumerate}
    \item when $0<b<\bar{b}$, two optimal allocations, $(n_{S1}^*,e_{S1}^*)$ and $(n_{S2}^*,e_{S2}^*)$, satisfy the household intertemporal maximization problem;
    \item when $b=\bar{b}$, $(n_{S1}^*,e_{S1}^*)$ and $(n_{S2}^*,e_{S2}^*)$ collapse into one double root;
    \item when $b>\bar{b}$, there is no economically meaningful optimal allocation.
\end{enumerate}
For each pair $(n_{S1}^*,e_{S1}^*)$ and $(n_{S2}^*,e_{S2}^*)$, the dynamic system (\ref{Dynam_Sys_subs_b}) admits a unique economically meaningful non-trivial equilibrium solution
$P_{S1}=(\bar{k}_{S1},\bar{N}_{S1})$ and $P_{S2}=(\bar{k}_{S2},\bar{N}_{S2})$.

\bigskip

\begin{proof}
See Appendix A.5.
\end{proof}

\end{proposition}

The proposition establishes the existence of equilibrium, but there is an important subtlety in how it is obtained. For small values of the subsidy ($b$), the household FOCs admit two distinct solutions for fertility and education investment. When these are substituted into the dynamic system, they give rise to two equilibrium points. This multiplicity does not reflect a property of the dynamic system itself, but rather the fact that two different optimal household plans are consistent with the same set of structural conditions. As $b$ increases, the two solutions approach one another and eventually collapse into a single allocation. Beyond that point, no interior solution to the household problem exists, and the dynamic system admits no equilibrium. Such a result provides an early indication of the risks associated with high subsidy rates, and suggests that intermediate values of $b$ may also give rise to complications. Both cases are examined in more detail in our numerical experiments later on.

\begin{proposition} \label{prop 6}
For each optimal allocation $(n_{Si}^*,e_{Si}^*)$, $i=1,2$, the corresponding
unique economically meaningful non-trivial equilibrium solution
$P_{Si}=(\bar{k}_{Si},\bar{N}_{Si})$ of the distorted dynamic system
(\ref{Dynam_Sys_subs_b}) is locally stable provided that:
\begin{align}
1-\frac{\alpha e_i^{\ast\alpha-1}\frac{\partial e_i^\ast}{\partial\bar{N}}\bar{k}^{\beta}
\frac{\partial n_i^\ast}{\partial\bar{k}}\bar{N}}{\alpha e_i^{\ast\alpha-1}\frac{\partial
e_i^\ast}{\partial\bar{k}}\bar{k}^{\beta}+\frac{\beta}{A}}-\frac
{1}{A\alpha e_i^{\ast\alpha-1}\frac{\partial e_i^\ast}{\partial\bar{k}}\bar{k}^{\beta}
+\beta}
&<-\frac{\partial n_i^\ast}{\partial\bar{N}}\bar{N}
<1-\frac
{A\alpha e_i^{\ast\alpha-1}\frac{\partial e_i^\ast}{\partial\bar{N}}\bar{k}^{\beta}
\frac{\partial n_i^\ast}{\partial\bar{k}}\bar{N}}{1+A\alpha e_i^{\ast\alpha-1}\frac{\partial
e_i^\ast}{\partial\bar{k}}\bar{k}^{\beta}+\beta}
\nonumber \\ \label{stab_subs_b} \\
\alpha e_i^{\ast\alpha-1}\frac{\partial e_i^\ast}{\partial\bar{k}}\bar{k}^{\beta}
+\frac{\beta}{A}
&>1 \nonumber
\end{align}
where
\begin{align*}
\frac{\partial n_i^\ast}{\partial \bar{k}}
&=
-\frac{F_{S,k}}{F_{S,n}}, \qquad \qquad
\frac{\partial n_i^\ast}{\partial \bar{N}}
=
-\frac{F_{S,N}}{F_{S,n}}, \\
\frac{\partial e_i^\ast}{\partial \bar{k}}
&=
\frac{\alpha}{1-\alpha}G_{S,k}, \qquad \qquad
\frac{\partial e_i^\ast}{\partial \bar{N}}
=
\frac{\alpha}{1-\alpha}G_{S,N},
\end{align*}
with
\begin{align*}
F_{S,n}
&=
4X^2n^*_i+\bar{k}\left[(\alpha-3)X+(1+\alpha)b\right],\\
F_{S,k}
&=
\frac{8X^2 {n^*_i}^2}{\bar{k}}+\left[3(\alpha-3)X+(1+\alpha)b\right]n^*_i+2(1-\alpha)\bar{k},\\
F_{S,N}
&=
\frac{\gamma X n^*_i}{\bar{N}}\left[4Xn^*_i+(\alpha-3)\bar{k}\right],
\end{align*}
and
\[
G_S(\bar{k},\bar{N})
=
\frac{\bar{k}(X-b)-X^2n_i^\ast}{\bar{k}-Xn_i^\ast},
\qquad X=\bar{k}^2\phi \bar{N}^\gamma,
\]
where
\begin{align*}
G_{S,k}
&=
\frac{
\left[
3X-b-\frac{4X^2n_i^\ast}{\bar{k}}-X^2\frac{\partial n_i^\ast}{\partial \bar{k}}
\right]
(\bar{k}-Xn_i^\ast)
-
\left[
\bar{k}(X-b)-X^2n_i^\ast
\right]
\left[
1-\frac{2Xn_i^\ast}{\bar{k}}-X\frac{\partial n_i^\ast}{\partial \bar{k}}
\right]
}{
(\bar{k}-Xn_i^\ast)^2
},\\[1em]
G_{S,N}
&=
\frac{
\left[
\frac{\gamma X}{\bar{N}}(\bar{k}-2Xn_i^\ast)-X^2\frac{\partial n_i^\ast}{\partial \bar{N}}
\right]
(\bar{k}-Xn_i^\ast)
+
\left[
\bar{k}(X-b)-X^2n_i^\ast
\right]
\left[
\frac{\gamma X n_i^\ast}{\bar{N}}+X\frac{\partial n_i^\ast}{\partial \bar{N}}
\right]
}{
(\bar{k}-Xn_i^\ast)^2
}.
\end{align*}

\bigskip

\begin{proof}
See Appendix A.6.
\end{proof}

\end{proposition}
Despite its apparent algebraic complexity, the local stability of the fixed point has an analogous interpretation to Proposition 4. We continue to have a corridor of stability for the fertility elasticity with respect to the population size, and education needs to respond to the size of aggregate human capital.

\subsection{Perfect complements ($\rho=-\infty$)}

Turning to the case when parenting time is complementary, the time constraint implies:
\[
h_t = 1 - 2 k_t n_t \phi N_t^\gamma.
\]
so that output becomes:
\begin{align}
y_t &= h_t k_t \nonumber \\ &= k_t - 2 k_t^2 \phi N_t^\gamma n_t.
\label{y_comp_dist}
\end{align}

Making use of the auxiliary variable $X_t \equiv k_t^2 \phi N_t^\gamma$, we have $y_t = k_t - 2 X_t n_t$ and
$\partial y_t/\partial n_t = -2 X_t$. Using Eq. (\ref{BC_dist}), consumption is therefore:
\[
c_t = (1-\tau)(k_t - 2X_t n_t) + B(n_t) - n_t e_t.
\]

The FOCs of the household intertemporal optimization problem become:
\begin{align}
    -\frac{n_t}{c_t} + \frac{\alpha}{e_t} = 0 \nonumber\\ \label{FOC_e_comp} \\
    \frac{1}{c_t}\left[-2(1-\tau)X_t + B'(n_t) - e_t \right] + \frac{1}{n_t} = 0 \nonumber
\end{align}

Combining both expressions in (\ref{FOC_e_comp}) and making use of the budget constraint (\ref{BC_dist}), the optimal fertility choice and education spending are:
\begin{align}
    n_t^\ast
=
\frac{(1-\alpha)\big[(1-\tau)k_t + B(n_t^\ast)\big]}
{4(1-\tau)X_t - (1+\alpha)B'(n_t^\ast)} \nonumber \\ \label{n_star_comp_dist}\\
e_t^\ast
=
\frac{\alpha}{1-\alpha}
\big[2(1-\tau)X_t - B'(n_t^\ast)\big] \nonumber
\end{align}
The first-order conditions under perfect complementarity deliver expressions that are structurally similar to those obtained in the substitutes case. The key difference lies in the implied allocations. Optimal fertility is half that of the previous case, while education investment is higher, though by less than a factor of two.

\subsubsection{Endogenizing the tax rate}

Recalling the government budget constraint $b n_t = \tau_t y_t$ and that $y_t = k_t - 2X_t n_t$, we can solve for the tax rate:
\begin{equation}
\tau_t
=
\frac{b n_t}{k_t - 2X_t n_t}.
\label{tau_comp_endog}
\end{equation}

Substituting Eq. (\ref{tau_comp_endog}) into the fertility optimal condition  (\ref{n_star_comp_dist}) yields:
\begin{equation}
8X_t^2 n_t^2
+
k_t\big[2(\alpha-3)X_t+(1+\alpha)b\big]n_t
+
(1-\alpha)k_t^2
=0.
\label{n_quad_comp}
\end{equation}
and the economically meaningful solution satisfies
$n_t\ge 0$ and $k_t - X_t n_t>0$.

\subsubsection{Dynamic system with endogenous taxation}

Equation (\ref{n_quad_comp}) also yields two real roots. Denoting them by
$n_{1,t}^{\ast}$ and $n_{2,t}^{\ast}$, we have:
\begin{equation*}
n_{i,t}^{\ast}
=
\frac{
k_t\left[2(3-\alpha)X_t-(1+\alpha)b \pm \sqrt{\Delta_t^{c}}\right]
}{
16X_t^2
},
\qquad i=1,2
\end{equation*}
where
\begin{equation*}
\Delta_t^{c}
=
\left[2(3-\alpha)X_t-(1+\alpha)b\right]^2
-
32(1-\alpha)X_t^2.
\end{equation*}
Thus, there are up to two economically meaningful fertility choices, provided
that $\Delta_t^{c}\geq 0$, $n_{i,t}^{\ast}\geq 0$, and
$k_t-2X_t n_{i,t}^{\ast}>0$.

Let $n_{i,t}^\ast(k_t,N_t)$ denote the feasible root of
(\ref{n_quad_comp}). Substituting Eq. (\ref{tau_subs_endog}) into the education investment optimal condition (\ref{n_star_comp_dist}), it follows:
\[
e_{i,t}^\ast
=
\frac{\alpha}{1-\alpha}
\left(
\frac{k_t(2X_t-b)-(2X_t)^2 n_{i,t}^\ast}
{k_t - 2X_t n_{i,t}^\ast}
\right).
\]

Thus, our two-dimensional dynamic system with distorting taxes becomes:
\begin{align}
k_{t+1}
&=
A \left[
\frac{\alpha}{1-\alpha}
\left(
\frac{k_t(2X_t-b)-(2X_t)^2 n_{i,t}^\ast}
{k_t - 2X_t n_{i,t}^\ast}
\right)
\right]^\alpha
k_t^\beta \nonumber \\
\label{Dynam_Sys_comp_b}
\\
N_{t+1}
&=
n_{i,t}^\ast N_t.
\nonumber
\end{align}
We proceed by stating and proving the following Proposition regarding the existence of a non-trivial equilibrium solution.

\begin{proposition} \label{prop 7}
When paternal and maternal parenting time are perfect complements, and the pro-natal subsidy is linear, $B(n_t)=bn_t$, there exists a threshold value $\bar{b}>0$ for which:
\begin{enumerate}
    \item when $0<b<\bar{b}$, two optimal allocations, $(n_{C1}^*,e_{C1}^*)$ and $(n_{C2}^*,e_{C2}^*)$, satisfy the household intertemporal maximization problem;
    \item when $b=\bar{b}$, $(n_{C1}^*,e_{C1}^*)$ and $(n_{C2}^*,e_{C2}^*)$ collapse into one double root;
    \item when $b>\bar{b}$, there is no economically meaningful optimal allocation.
\end{enumerate}
For each pair $(n_{C1}^*,e_{C1}^*)$ and $(n_{C2}^*,e_{C2}^*)$, the dynamic system (\ref{Dynam_Sys_comp_b}) admits a unique economically meaningful non-trivial equilibrium solution
$P_{C1}=(\bar{k}_{C1},\bar{N}_{C1})$ and $P_{C2}=(\bar{k}_{C2},\bar{N}_{C2})$, such that:
\begin{equation*}
\bar{k}_{Ci}=\bar{k}_{Si} \qquad \qquad
\bar{N}_{Ci}=\Omega\bar{N}_{Si}
\end{equation*}
where
\begin{align*}
\Omega &=\frac{1}{2^{1/\gamma}}\in(0,1) \\
i &=1,2
\end{align*}

\bigskip

\begin{proof}
See Appendix A.7.
\end{proof}

\end{proposition}

We are able to confirm our previous result that, in equilibrium, the steady-state level of human capital does not depend on the degree of substitutability between parental time inputs. Although education investment is lower under perfect substitutability, decreasing returns to skill accumulation equalize human capital across the two regimes in the long run. The same does not hold for population size. Complementarity in parental time leads to a smaller steady-state population through a mechanism mediated by land availability. In terms of the local stability properties of the equilibrium point, we state and prove the following Proposition.

\begin{proposition} \label{prop 8}
For each optimal allocation $(n_{Ci}^*,e_{Ci}^*)$, $i=1,2$, the corresponding
unique economically meaningful non-trivial equilibrium solution
$P_{Ci}=(\bar{k}_{Ci},\bar{N}_{Ci})$ of the distorted dynamic system
(\ref{Dynam_Sys_comp_b}) is locally stable provided that (\ref{stab_subs_b}) is
satisfied, and we further have:
\begin{align*}
\frac{\partial n_i^\ast}{\partial \bar{k}}
&=
-\frac{F_{C,k}}{F_{C,n}}, \qquad \qquad
\frac{\partial n_i^\ast}{\partial \bar{N}}
=
-\frac{F_{C,N}}{F_{C,n}}, \\
\frac{\partial e_i^\ast}{\partial \bar{k}}
&=
\frac{\alpha}{1-\alpha}G_{C,k}, \qquad \qquad
\frac{\partial e_i^\ast}{\partial \bar{N}}
=
\frac{\alpha}{1-\alpha}G_{C,N},
\end{align*}
with
\begin{align*}
F_{C,n}
&=
16X^2n^*_i+\bar{k}\left[2(\alpha-3)X+(1+\alpha)b\right],\\
F_{C,k}
&=
\frac{32X^2{n^*_i}^2}{\bar{k}}+\left[6(\alpha-3)X+(1+\alpha)b\right]n^*_i+2(1-\alpha)\bar{k},\\
F_{C,N}
&=
\frac{2\gamma X n^*_i}{\bar{N}}\left[8Xn^*_i+(\alpha-3)\bar{k}\right],
\end{align*}
and
\[
G_C(\bar{k},\bar{N})
=
\frac{\bar{k}(2X-b)-4X^2n_i^\ast}{\bar{k}-2Xn_i^\ast},
\qquad X=\bar{k}^2\phi \bar{N}^\gamma,
\]
where
\begin{align*}
G_{C,k}
&=
\frac{
\left[
6X-b-\frac{16X^2n_i^\ast}{\bar{k}}-4X^2\frac{\partial n_i^\ast}{\partial \bar{k}}
\right]
(\bar{k}-2Xn_i^\ast)
-
\left[
\bar{k}(2X-b)-4X^2n_i^\ast
\right]
\left[
1-\frac{4Xn_i^\ast}{\bar{k}}-2X\frac{\partial n_i^\ast}{\partial \bar{k}}
\right]
}{
(\bar{k}-2Xn_i^\ast)^2
},\\[1em]
G_{C,N}
&=
\frac{
\left[
\frac{2\gamma X}{\bar{N}}(\bar{k}-4Xn_i^\ast)-4X^2\frac{\partial n_i^\ast}{\partial \bar{N}}
\right]
(\bar{k}-2Xn_i^\ast)
+
\left[
\bar{k}(2X-b)-4X^2n_i^\ast
\right]
\left[
\frac{2\gamma X n_i^\ast}{\bar{N}}+2X\frac{\partial n_i^\ast}{\partial \bar{N}}
\right]
}{
(\bar{k}-2Xn_i^\ast)^2
}.
\end{align*}

\bigskip

\begin{proof}
See Appendix A.8.
\end{proof}

\end{proposition}

\subsection{Numerical illustrations}

\begin{figure}[tbp]
  \centering
  \subfloat[]{\includegraphics[width=2.5in]{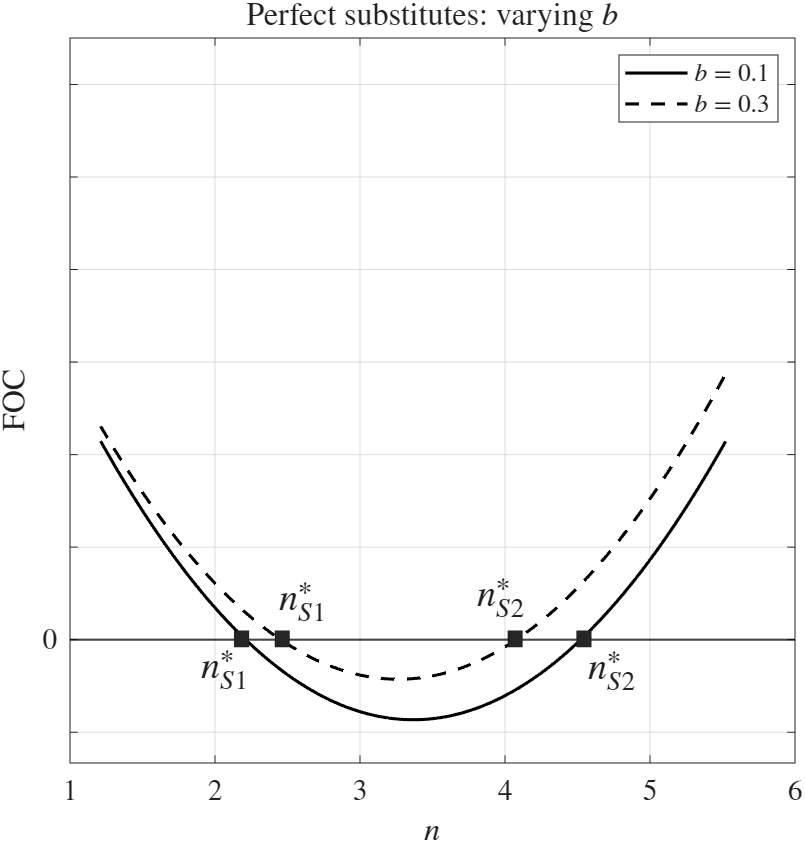}} \hspace{0.75cm}
  \subfloat[]{\includegraphics[width=2.5in]{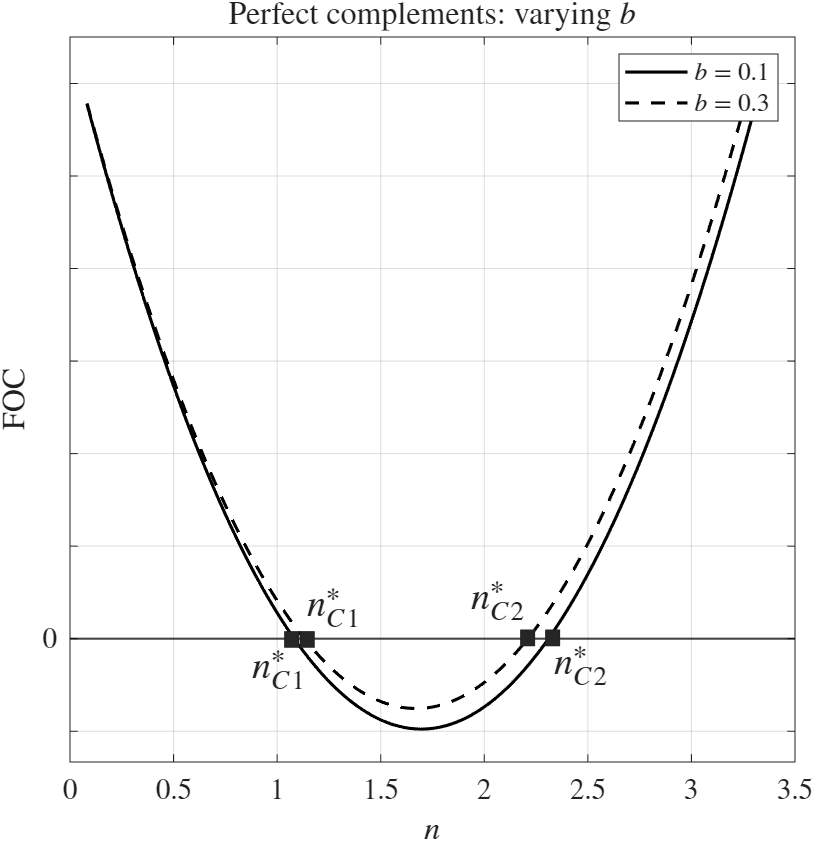}} \\
  \subfloat[]{\includegraphics[width=2.5in]{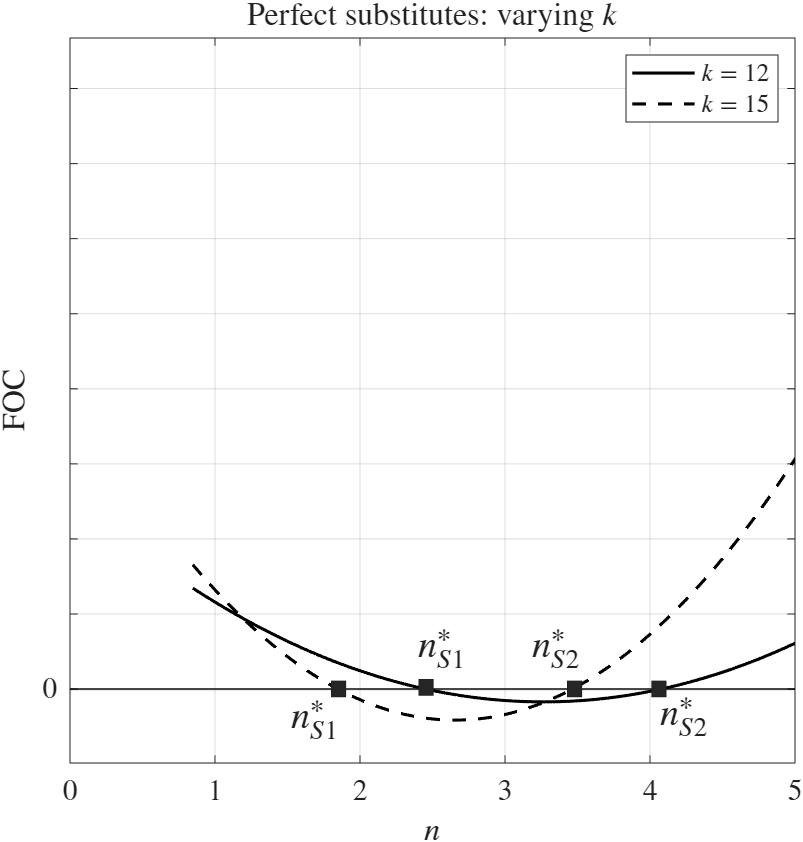}} \hspace{0.75cm}
  \subfloat[]{\includegraphics[width=2.5in]{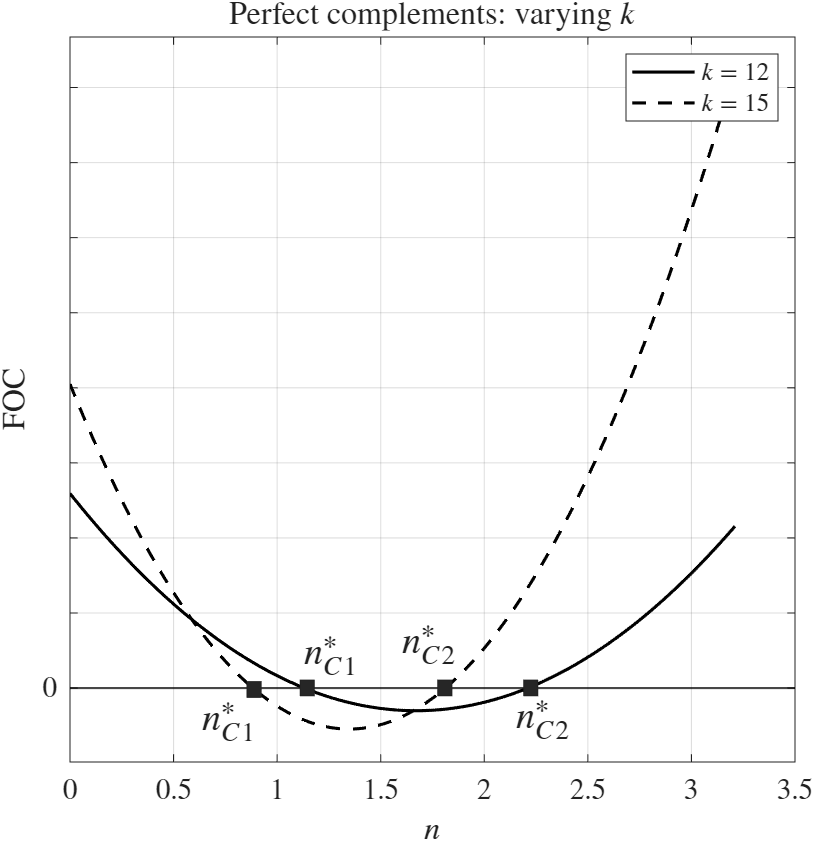}} \\
  \subfloat[]{\includegraphics[width=2.5in]{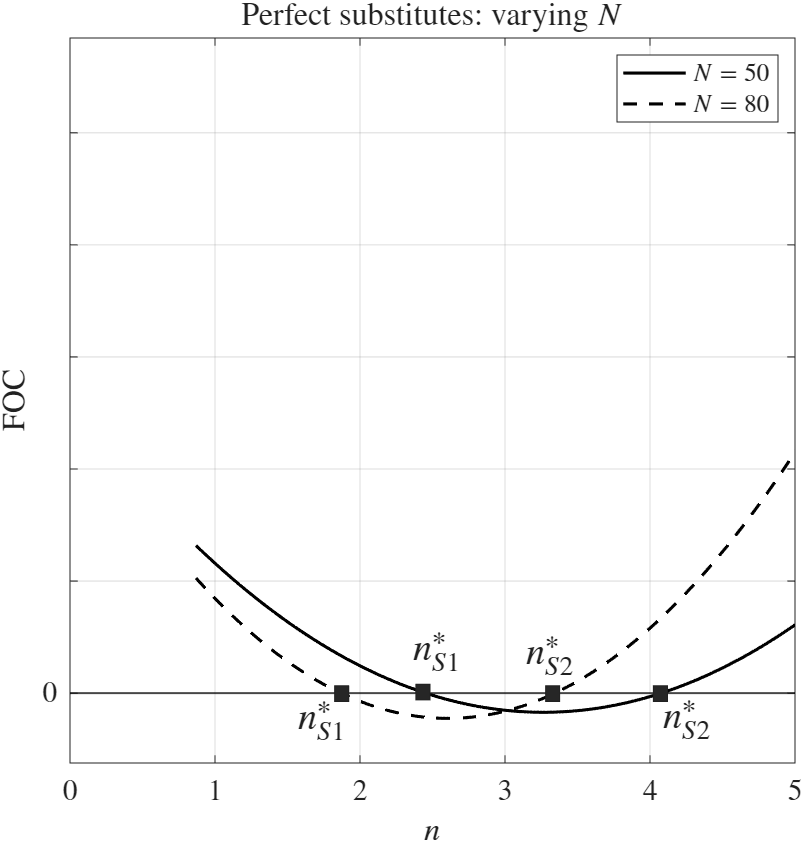}} \hspace{0.75cm}
  \subfloat[]{\includegraphics[width=2.5in]{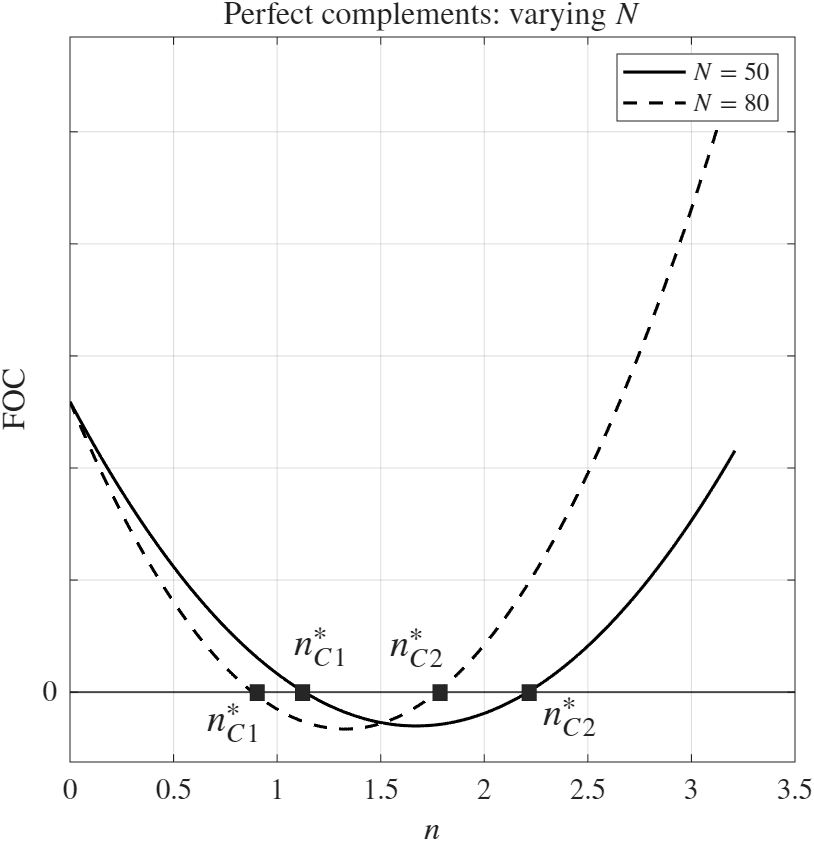}} \\
  \caption{Comparing equilibria for different education returns. Parameters $A=3.75$, $\beta=0.5$, $\gamma=0.5$, $\phi= 0.0025$.}\label{FOC}
\end{figure}

Similar to what we did with our baseline model, we provide a numerical example to compare the two main cases studied here. We use \cite{delacroix_gosseries_2012} and South Korean data as a loose reference for the exercise, which should be interpreted only qualitatively for illustrative purposes. Fig. \ref{FOC} confirms that the FOCs are U-shaped in $n$ so that there are two optimal possible allocations: one with low ($n^*_1$) and the other with high ($n^*_2$) fertility. Notice that across cases, perfect substitution delivers significantly higher fertility rates, and often $n^*_{S1}$ is higher than $n^*_{C2}$.

The quadratic relationship results from two antagonizing forces that arise precisely because households do not internalize the government budget constraint in their intertemporal optimization. On one hand, a higher tax rate $\tau$, required to balance the public budget ex post, reduces household disposable 
income, leading families to adjust their fertility downward. On the other hand, a higher $\tau$ finances a more generous subsidy per child, making additional children less costly and pushing fertility upward. The net effect depends on which force dominates, and, in turn, that depends on where the economy lies along the fertility distribution. When households are at the low-fertility optimal allocation, the marginal utility of an additional child is relatively high, and the subsidy effect prevails, successfully raising fertility. However, when households are at the high-fertility allocation, the marginal utility of another child is low, and the income effect of the higher tax dominates. In this case, it is optimal for households to reduce the number of their children in response to the subsidy. The quadratic structure of the first-order condition is thus a direct reflection of this tension between the fiscal cost and the direct incentive embedded in the pro-natalist policy.

We show in panels (a) and (b) how optimal fertility allocations respond to an increase in the pro-natalist subsidy. The direction of the effect is not uniform. Whether the policy raises or reduces fertility depends on whether households find themselves at $n^*_1$ or $n^*_2$. The subsidy achieves its intended purpose only when the economy operates at the lowest point. Under perfect complementarity, panel (b) shows that households are nearly unresponsive to the policy. In fact, fertility barely moves despite a substantial increase in $b$ from 0.1 to 0.3.

Panels (c) and (d) report the response of optimal fertility to a change in aggregate human capital, which is equivalent to tracing the optimal reaction to an exogenous increase in the human capital endowment of the incoming adult generation. In both regimes, the effect is a strong reduction in fertility. The mechanism is straightforward. Higher human capital raises the opportunity cost of time spent outside the labor market, so that a higher $k$ leads households to optimally choose fewer children. Panels (e) and (f) document a similar negative relationship between population size and fertility, though the underlying channel is distinct. A larger population reduces the land available to support additional children, tightening the space constraint faced by households. This shifts the first-order condition parabola to the left, implying lower optimal fertility at any given level of human capital.

\begin{figure}[tbp]
  \centering
  \subfloat[]{\includegraphics[width=5in]{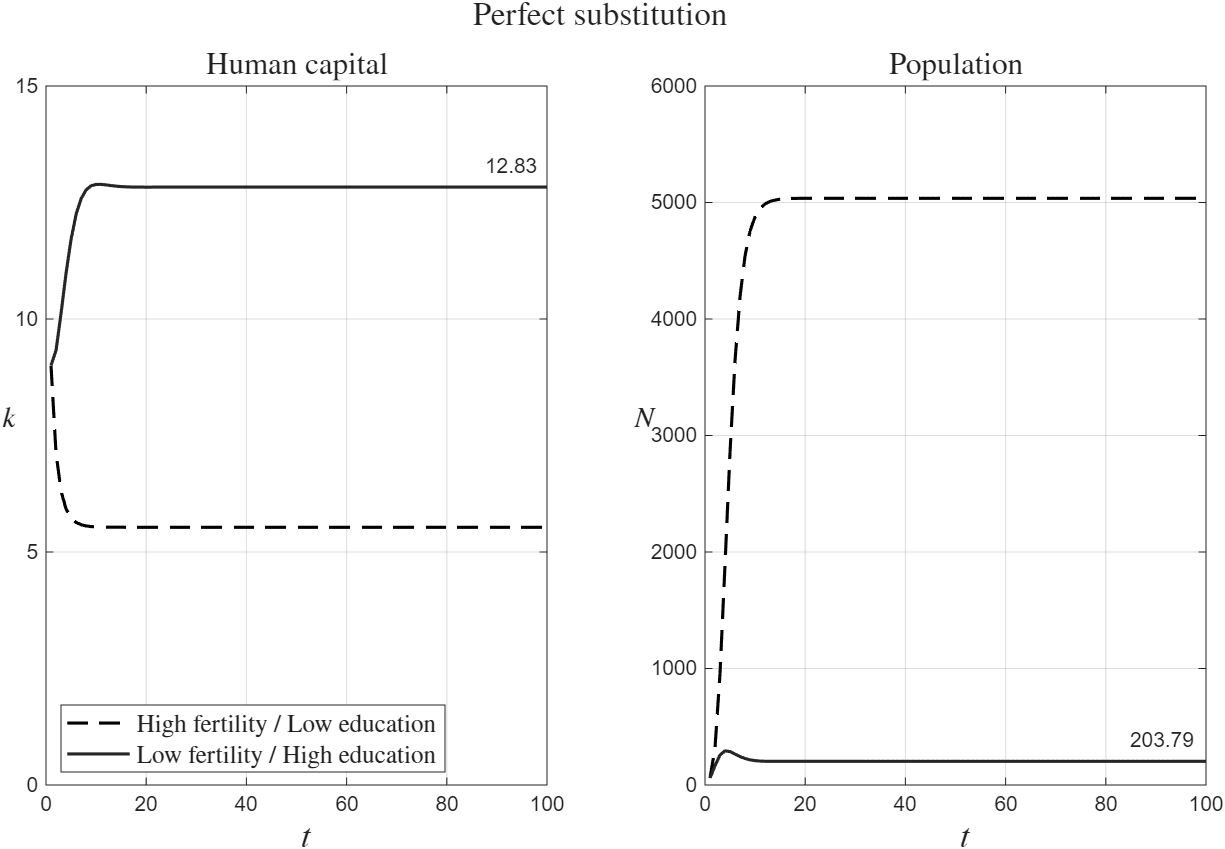}} \\
  \subfloat[]{\includegraphics[width=5in]{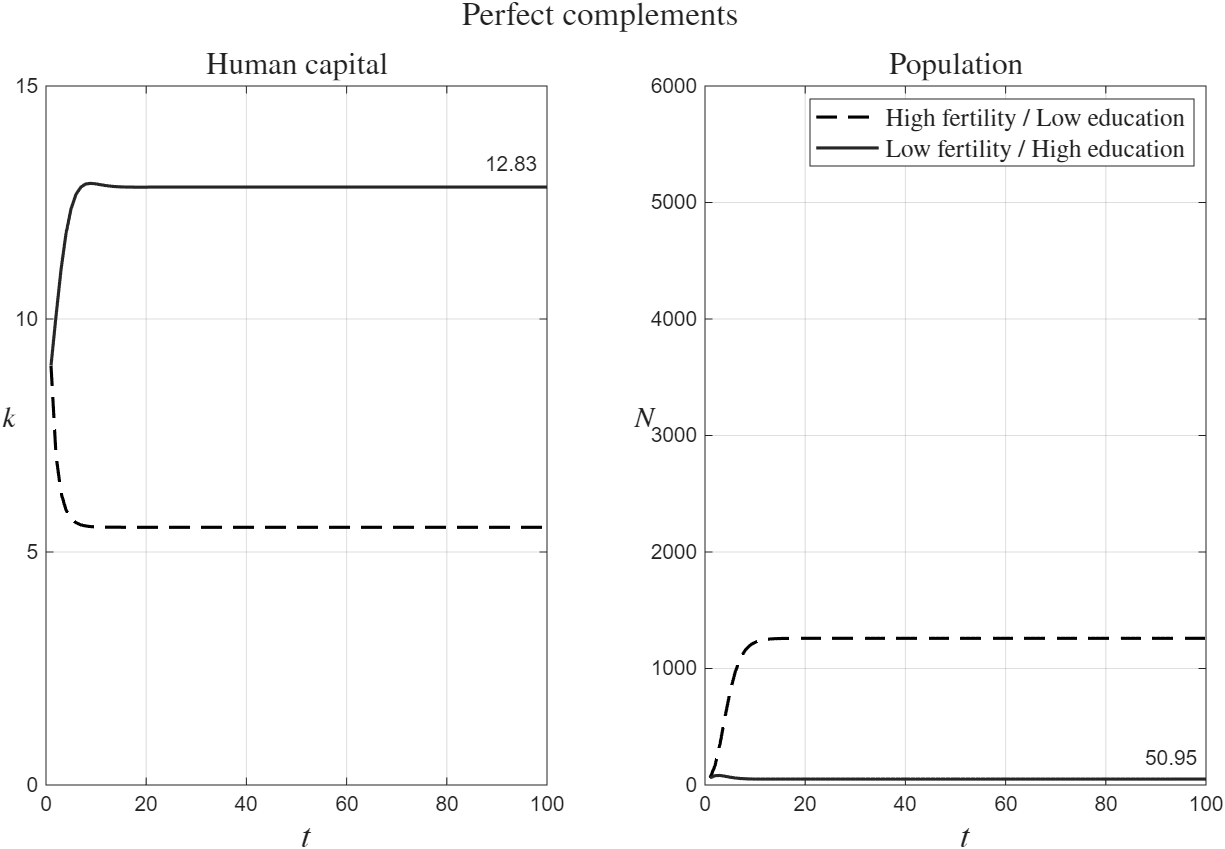}}
  \caption{Comparing equilibrium points of human capital and population size. Parameters $A=3.75$, $\beta=0.5$, $\gamma=0.5$, $\phi= 0.0025$, $b=0.1$. In the perfect complements case, $\bar{k}$ is 12.83 and $\bar{N}$ is 50.95, close to Korea's current levels, 12 and 51.6, respectively.}\label{Timeseries}
\end{figure}

We conclude this exercise by reporting the time series of human capital 
and population under perfect substitutability and perfect complementarity. Fig.~\ref{Timeseries} illustrates the quantity-quality trade-off in children. Continuous lines correspond to the case in which households invest more heavily in education, accepting lower fertility as a result. Dotted lines represent the opposite case, with substantially higher population but considerably lower human capital. The comparison of panels (a) and (b) brings us back to the central message of the paper. Complementarity in intra-household childcare allocation necessarily leads to a smaller population, with no corresponding impact on aggregate human capital.

\section{Final considerations}

There has long been an apparent consensus in the literature on intra-household allocation and fertility that greater paternal involvement in childcare alleviates maternal time constraints, enabling mothers to increase their labor supply or leisure. The Korean evidence challenges that view. Increases in fathers' childcare time have coincided not with a reduction, but with a further increase, in the time mothers allocate to caring for their children.

This paper developed an OLG growth model to study a mechanism that at least partially explains the puzzle. The core argument rests on the structure of the childcare technology. When parental time inputs are substitutes, the conventional result holds: greater paternal involvement eases the maternal time burden. When they are complements, the opposite occurs, and increased paternal 
involvement raises maternal childcare time rather than reducing it. The elasticity of substitution between parental time inputs should not be interpreted purely as a preference parameter. It also reflects the set of skills that mothers and fathers bring to child-rearing. High substitutability captures a situation in which both parents are capable of performing the same household tasks interchangeably. As the elasticity falls, parents become increasingly specialized, with each 
unable to replicate what the other does. The simple example of cooking and driving illustrates the point. Two parents who can both cook and drive are substitutes; a father who can cook but not drive, paired with a mother who can drive but not cook, become complements.

The model was further used to assess the effectiveness of fertility 
subsidies under two alternative fiscal arrangements. In the first, the 
government maintains a balanced budget at all times, and households 
fully internalize the implied tax burden in their optimization problem in the form of Ricardian equivalence by design. In the second, the 
budget is balanced ex post, after households have made their fertility 
and education decisions, so that the subsidy has a distortionary effect 
on household choices. It was shown that pro-natalist subsidies may generate an unintended anti-fertility bias, working against their stated purpose. Numerical experiments confirmed that the model is consistent with 
additional stylized facts, including the quantity-quality trade-off 
between fertility and investment in human capital.

Several avenues remain open for future research. The framework 
developed here deliberately abstracts from the literature on domestic 
labor and intra-household bargaining. This was done to prioritize parsimony to isolate the mechanism of interest. Of course, this need not be a permanent restriction. The childcare technology introduced in this paper could serve as a natural building block for models that explicitly incorporate asymmetric bargaining power between partners, or that allow for the possibility of divorce. Integrating these elements would enrich the framework and shed further light on how the distribution of power and 
risk within the household shapes fertility and human capital outcomes.

\FloatBarrier

\newpage

\appendix

\numberwithin{equation}{section}
\numberwithin{figure}{section}

\section{Mathematical Appendix} 

\subsection{Proof of Proposition 1} 

When the time dad and mom dedicate to parenting are substitutes, from the
equilibrium conditions, we obtain the following isoquants:%
\begin{align}
\bar{k} &  =A^{\frac{1}{1-\beta-2\alpha}}\left[  \left(  \frac{\alpha
}{1-\alpha}\right)  \phi\bar{N}^{\gamma}\right]  ^{\frac{\alpha}%
{1-\beta-2\alpha}}\label{iso1_subs}\\
\bar{k} &  =\left(  \frac{1-\alpha}{2\phi\bar{N}^{\gamma}}\right)
\label{iso2_subs}%
\end{align}
Combining (\ref{iso1_subs}) and (\ref{iso2_subs}), we have:%
\[
\frac{1-\alpha}{2\phi\bar{N}^{\gamma}} =A^{\frac{1}%
{1-\beta-2\alpha}}\left[  \left(  \frac{\alpha}{1-\alpha}\right)  \phi\bar
{N}^{\gamma}\right]  ^{\frac{\alpha}{1-\beta-2\alpha}}%
\]
Isolating $\bar{N}$, it follows:%
\begin{align*}
 \frac{1-\alpha}{2\phi}    & =A^{\frac{1}{1-\beta-2\alpha}%
}\left[  \left(  \frac{\alpha}{1-\alpha}\right)  \phi\right]  ^{\frac{\alpha
}{1-\beta-2\alpha}}\bar{N}^{\gamma+\frac{\alpha\gamma}{1-\beta-2\alpha}}\\
\left(  \frac{1-\alpha}{2\phi}\right)  A^{\frac{-1}{1-\beta-2\alpha}}\left[
\left(  \frac{\alpha}{1-\alpha}\right)  \phi\right]  ^{\frac{-\alpha}%
{1-\beta-2\alpha}}  & =\bar{N}^{\frac{\left(  1-\beta-\alpha\right)  \gamma
}{1-\beta-2\alpha}}\\
\bar{N}^{\left(  1-\beta-\alpha\right)  \gamma}  & =\left(  \frac{1-\alpha
}{2\phi}\right)  ^{1-\beta-2\alpha}A^{-1}\left[  \left(  \frac{\alpha
}{1-\alpha}\right)  \phi\right]  ^{-\alpha}\\
\bar{N}  & =A^{\frac{-1}{\left(  1-\beta-\alpha\right)  \gamma}}\left(
\frac{1-\alpha}{2\phi}\right)  ^{\frac{1-\beta-2\alpha}{\left(  1-\beta
-\alpha\right)  \gamma}}\left[  \left(  \frac{\alpha}{1-\alpha}\right)
\phi\right]  ^{\frac{-\alpha}{\left(  1-\beta-\alpha\right)  \gamma}}%
\end{align*}

\bigskip Looking at the optimal solutions of $n$ and $e$, notice that we can
disaggregate between:%
\begin{equation}
\bar{N}=A^{\frac{-1}{\left(  1-\beta-\alpha\right)  \gamma}}\underset
{\text{{\scriptsize Fertility}}}{\underbrace{\left(  \frac{1-\alpha}{2\phi
}\right)  }}^{\frac{1-\beta-2\alpha}{\left(  1-\beta-\alpha\right)  \gamma}%
}\underset{\text{{\scriptsize Education}}}{\underbrace{\left[  \left(
\frac{\alpha}{1-\alpha}\right)  \phi\right]  }}^{\frac{-\alpha}{\left(
1-\beta-\alpha\right)  \gamma}}\label{eq_N_subs}%
\end{equation}
Finally, substituting (\ref{eq_N_subs}) into (\ref{iso2_subs}), we have:
\begin{align}
\bar{k}  & =\left(  \frac{1-\alpha}{2\phi}\right)  A^{\frac{\gamma}{\left(
1-\beta-\alpha\right)  \gamma}}\left(  \frac{1-\alpha}{2\phi}\right)
^{\frac{-\left(  1-\beta-2\alpha\right)  \gamma}{\left(  1-\beta
-\alpha\right)  \gamma}}\left[  \left(  \frac{\alpha}{1-\alpha}\right)
\phi\right]  ^{\frac{\alpha\gamma}{\left(  1-\beta-\alpha\right)  \gamma}%
}\nonumber\\
& =\left(  \frac{1-\alpha}{2\phi}\right)  A^{\frac{1}{1-\beta-\alpha}}\left(
\frac{1-\alpha}{2\phi}\right)  ^{\frac{-1+\beta+2\alpha}{1-\beta-\alpha}%
}\left[  \left(  \frac{\alpha}{1-\alpha}\right)  \phi\right]  ^{\frac{\alpha
}{1-\beta-\alpha}}\nonumber\\
& =A^{\frac{1}{1-\beta-\alpha}}\left(  \frac{1-\alpha}{2\phi}\right)
^{\frac{\alpha}{1-\beta-\alpha}}\left[  \left(  \frac{\alpha}{1-\alpha
}\right)  \phi\right]  ^{\frac{\alpha}{1-\beta-\alpha}}\nonumber\\
& =A^{\frac{1}{1-\beta-\alpha}}\left[  \left(  \frac{\alpha}{2\phi}\right)
\phi\right]  ^{\frac{\alpha}{1-\beta-\alpha}}\nonumber\\
& =A^{\frac{1}{1-\beta-\alpha}}\left(  \frac{\alpha}{2}\right)  ^{\frac
{\alpha}{1-\beta-\alpha}}\nonumber\\
& =\left(  A^{\frac{1}{\alpha}}\frac{\alpha}{2}\right)  ^{\frac{\alpha
}{1-\beta-\alpha}}\nonumber\\
& =\left(  A^{\frac{1}{\alpha}}\frac{\alpha}{2}\right)  ^{\frac{\alpha\gamma
}{\left(  1-\beta-\alpha\right)  \gamma}}\label{eq_k_subs}%
\end{align}

\subsection{Proof of Proposition 2} 

Recall our dynamic system:%
\begin{align}
k_{t+1}  &  =Ae^{\ast\alpha}\left(  k_{t},N_{t}\right)  k_{t}^{\beta
}\nonumber\\
& \label{Dynam_Sys_base}\\
N_{t+1}  &  =n^{\ast}\left(  k_{t},N_{t}\right)  N_{t}\nonumber
\end{align}
where $e^{\ast}\left(  k_{t},N_{t}\right)  $ and $n^{\ast}\left(  k_{t}%
,N_{t}\right)  $ are optimal education investment and fertility rates obtained
from the household intertemporal optimization problem. In steady-state,
$k_{t}=k_{t+1}=\bar{k}$ and $N_{t}=N_{t+1}=\bar{N}$, so it follows that:%
\begin{align}
e^{\ast\alpha}\left(  k_{t},N_{t}\right)   &  =\frac{\bar{k}^{1-\beta}}%
{A}\nonumber\\
& \label{Auxiliary}\\
n^{\ast}\left(  k_{t},N_{t}\right)   &  =1\nonumber
\end{align}

The Jacobian matrix is given by:%
\[
J=\left[
\begin{array}
[c]{cc}%
j_{11} & j_{12}\\
j_{21} & j_{22}%
\end{array}
\right]
\]
where%
\begin{align*}
j_{11}  &  =A\left(  \alpha\frac{\partial e^{\ast\alpha-1}}{\partial\bar{k}%
}\bar{k}^{\beta}+\beta e^{\ast\alpha}\left(  \bar{k},\bar{N}\right)  \bar
{k}^{\beta-1}\right)  >0\\
j_{12}  &  =A\alpha\frac{\partial e^{\ast\alpha-1}}{\partial\bar{N}}\bar
{k}^{\beta}>0\\
j_{21}  &  =\frac{\partial n^{\ast}}{\partial\bar{k}}\bar{N}<0\\
j_{22}  &  =\frac{\partial n^{\ast}}{\partial\bar{N}}\bar{N}+n^{\ast}\left(
\bar{k},\bar{N}\right)  \gtreqqless0
\end{align*}

Thus:%
\begin{align*}
\text{tr}J &  =A\left(  \alpha\frac{\partial e^{\ast\alpha-1}}{\partial\bar
{k}}\bar{k}^{\beta}+\beta e^{\ast\alpha}\left(  \bar{k},\bar{N}\right)
\bar{k}^{\beta-1}\right)  +\frac{\partial n^{\ast}}{\partial\bar{N}}\bar
{N}+n^{\ast}\left(  \bar{k},\bar{N}\right)  \gtreqqless0\\
\det J &  =A\left(  \alpha\frac{\partial e^{\ast\alpha-1}}{\partial\bar{k}%
}\bar{k}^{\beta}+\beta e^{\ast\alpha}\left(  \bar{k},\bar{N}\right)  \bar
{k}^{\beta-1}\right)  \left(  \frac{\partial n^{\ast}}{\partial\bar{N}}\bar
{N}+n^{\ast}\left(  \bar{k},\bar{N}\right)  \right)  -A\alpha\frac{\partial
e^{\ast\alpha-1}}{\partial\bar{N}}\bar{k}^{\beta}\frac{\partial n^{\ast}%
}{\partial\bar{k}}\bar{N}\gtreqqless0
\end{align*}
Making use of (\ref{Auxiliary}), the local stability conditions become (see \citealp{MedioLines2001}):%
\begin{align*}
&  \text{(i) }1+\text{tr}J+\det J\\
&  =1+A\left(  \alpha\frac{\partial e^{\ast\alpha-1}}{\partial\bar{k}}\bar
{k}^{\beta}+\beta e^{\ast\alpha}\left(  \bar{k},\bar{N}\right)  \bar{k}%
^{\beta-1}\right)  +\frac{\partial n^{\ast}}{\partial\bar{N}}\bar{N}+n^{\ast
}\left(  \bar{k},\bar{N}\right)  \\
&  +A\left(  \alpha\frac{\partial e^{\ast\alpha-1}}{\partial\bar{k}}\bar
{k}^{\beta}+\beta e^{\ast\alpha}\left(  \bar{k},\bar{N}\right)  \bar{k}%
^{\beta-1}\right)  \left(  \frac{\partial n^{\ast}}{\partial\bar{N}}\bar
{N}+n^{\ast}\left(  \bar{k},\bar{N}\right)  \right)  -A\alpha\frac{\partial
e^{\ast\alpha-1}}{\partial\bar{N}}\bar{k}^{\beta}\frac{\partial n^{\ast}%
}{\partial\bar{k}}\bar{N}\\
&  =1+A\left(  \alpha\frac{\partial e^{\ast\alpha-1}}{\partial\bar{k}}\bar
{k}^{\beta}+\frac{\beta}{A}\right)  +\frac{\partial n^{\ast}}{\partial\bar{N}%
}\bar{N}+A\left(  \alpha\frac{\partial e^{\ast\alpha-1}}{\partial\bar{k}}%
\bar{k}^{\beta}+\frac{\beta}{A}\right)  \frac{\partial n^{\ast}}{\partial
\bar{N}}\bar{N}-A\alpha\frac{\partial e^{\ast\alpha-1}}{\partial\bar{N}}%
\bar{k}^{\beta}\frac{\partial n^{\ast}}{\partial\bar{k}}\bar{N}>0
\end{align*}%
\begin{align*}
1+A\left(  \alpha\frac{\partial e^{\ast\alpha-1}}{\partial\bar{k}}\bar
{k}^{\beta}+\frac{\beta}{A}\right)  -A\alpha\frac{\partial e^{\ast\alpha-1}%
}{\partial\bar{N}}\bar{k}^{\beta}\frac{\partial n^{\ast}}{\partial\bar{k}}%
\bar{N} &  >-\frac{\partial n^{\ast}}{\partial\bar{N}}\bar{N}-A\left(
\alpha\frac{\partial e^{\ast\alpha-1}}{\partial\bar{k}}\bar{k}^{\beta}%
+\frac{\beta}{A}\right)  \frac{\partial n^{\ast}}{\partial\bar{N}}\bar{N}\\
1+A\left(  \alpha\frac{\partial e^{\ast\alpha-1}}{\partial\bar{k}}\bar
{k}^{\beta}+\frac{\beta}{A}\right)  -A\alpha\frac{\partial e^{\ast\alpha-1}%
}{\partial\bar{N}}\bar{k}^{\beta}\frac{\partial n^{\ast}}{\partial\bar{k}}%
\bar{N} &  >-\left[  1+A\left(  \alpha\frac{\partial e^{\ast\alpha-1}%
}{\partial\bar{k}}\bar{k}^{\beta}+\frac{\beta}{A}\right)  \right]
\frac{\partial n^{\ast}}{\partial\bar{N}}\bar{N}\\
-\frac{\partial n^{\ast}}{\partial\bar{N}}\bar{N} &  <\frac{1+A\left(
\alpha\frac{\partial e^{\ast\alpha-1}}{\partial\bar{k}}\bar{k}^{\beta}%
+\frac{\beta}{A}\right)  -A\alpha\frac{\partial e^{\ast\alpha-1}}{\partial
\bar{N}}\bar{k}^{\beta}\frac{\partial n^{\ast}}{\partial\bar{k}}\bar{N}%
}{1+A\left(  \alpha\frac{\partial e^{\ast\alpha-1}}{\partial\bar{k}}\bar
{k}^{\beta}+\frac{\beta}{A}\right)  }\\
-\frac{\partial n^{\ast}}{\partial\bar{N}}\bar{N} &  <1-\frac{A\alpha
\frac{\partial e^{\ast\alpha-1}}{\partial\bar{N}}\bar{k}^{\beta}\frac{\partial
n^{\ast}}{\partial\bar{k}}\bar{N}}{1+A\left(  \alpha\frac{\partial
e^{\ast\alpha-1}}{\partial\bar{k}}\bar{k}^{\beta}+\frac{\beta}{A}\right)  }%
\end{align*}%
\begin{align*}
&  \text{(ii) }1-\text{tr}J+\det J\\
&  =1-A\left(  \alpha\frac{\partial e^{\ast\alpha-1}}{\partial\bar{k}}\bar
{k}^{\beta}+\beta e^{\ast\alpha}\left(  \bar{k},\bar{N}\right)  \bar{k}%
^{\beta-1}\right)  -\frac{\partial n^{\ast}}{\partial\bar{N}}\bar{N}-n^{\ast
}\left(  \bar{k},\bar{N}\right)  \\
&  +A\left(  \alpha\frac{\partial e^{\ast\alpha-1}}{\partial\bar{k}}\bar
{k}^{\beta}+\beta e^{\ast\alpha}\left(  \bar{k},\bar{N}\right)  \bar{k}%
^{\beta-1}\right)  \left(  \frac{\partial n^{\ast}}{\partial\bar{N}}\bar
{N}+n^{\ast}\left(  \bar{k},\bar{N}\right)  \right)  -A\alpha\frac{\partial
e^{\ast\alpha-1}}{\partial\bar{N}}\bar{k}^{\beta}\frac{\partial n^{\ast}%
}{\partial\bar{k}}\bar{N}\\
&  =1+A\left(  \alpha\frac{\partial e^{\ast\alpha-1}}{\partial\bar{k}}\bar
{k}^{\beta}+\beta e^{\ast\alpha}\left(  \bar{k},\bar{N}\right)  \bar{k}%
^{\beta-1}\right)  \left(  \frac{\partial n^{\ast}}{\partial\bar{N}}\bar
{N}+n^{\ast}\left(  \bar{k},\bar{N}\right)  -1\right)  \\
&  -\left(  \frac{\partial n^{\ast}}{\partial\bar{N}}\bar{N}+n^{\ast}\left(
\bar{k},\bar{N}\right)  \right)  -A\alpha\frac{\partial e^{\ast\alpha-1}%
}{\partial\bar{N}}\bar{k}^{\beta}\frac{\partial n^{\ast}}{\partial\bar{k}}%
\bar{N}\\
&  =1+A\left(  \alpha\frac{\partial e^{\ast\alpha-1}}{\partial\bar{k}}\bar
{k}^{\beta}+\frac{\beta}{A}\right)  \frac{\partial n^{\ast}}{\partial\bar{N}%
}\bar{N}-\left(  \frac{\partial n^{\ast}}{\partial\bar{N}}\bar{N}+1\right)
-A\alpha\frac{\partial e^{\ast\alpha-1}}{\partial\bar{N}}\bar{k}^{\beta}%
\frac{\partial n^{\ast}}{\partial\bar{k}}\bar{N}>0
\end{align*}%
\begin{align*}
A\left(  \alpha\frac{\partial e^{\ast\alpha-1}}{\partial\bar{k}}\bar{k}%
^{\beta}+\frac{\beta}{A}\right)  \frac{\partial n^{\ast}}{\partial\bar{N}}%
\bar{N}-\frac{\partial n^{\ast}}{\partial\bar{N}}\bar{N}-A\alpha\frac{\partial
e^{\ast\alpha-1}}{\partial\bar{N}}\bar{k}^{\beta}\frac{\partial n^{\ast}%
}{\partial\bar{k}}\bar{N} &  >0\\
\left[  A\left(  \alpha\frac{\partial e^{\ast\alpha-1}}{\partial\bar{k}}%
\bar{k}^{\beta}+\frac{\beta}{A}\right)  -1\right]  \frac{\partial n^{\ast}%
}{\partial\bar{N}}\bar{N}-A\alpha\frac{\partial e^{\ast\alpha-1}}{\partial
\bar{N}}\bar{k}^{\beta}\frac{\partial n^{\ast}}{\partial\bar{k}}\bar{N} &  >0
\end{align*}%
\begin{align*}
&  \text{(iii) }1-\det J\\
&  =1-A\left(  \alpha\frac{\partial e^{\ast\alpha-1}}{\partial\bar{k}}\bar
{k}^{\beta}+\beta e^{\ast\alpha}\left(  \bar{k},\bar{N}\right)  \bar{k}%
^{\beta-1}\right)  \left(  \frac{\partial n^{\ast}}{\partial\bar{N}}\bar
{N}+n^{\ast}\left(  \bar{k},\bar{N}\right)  \right)  +A\alpha\frac{\partial
e^{\ast\alpha-1}}{\partial\bar{N}}\bar{k}^{\beta}\frac{\partial n^{\ast}%
}{\partial\bar{k}}\bar{N}\\
&  =1-A\left(  \alpha\frac{\partial e^{\ast\alpha-1}}{\partial\bar{k}}\bar
{k}^{\beta}+\beta e^{\ast\alpha}\left(  \bar{k},\bar{N}\right)  \bar{k}%
^{\beta-1}\right)  \frac{\partial n^{\ast}}{\partial\bar{N}}\bar{N}\\
&  -A\left(  \alpha\frac{\partial e^{\ast\alpha-1}}{\partial\bar{k}}\bar
{k}^{\beta}+\beta e^{\ast\alpha}\left(  \bar{k},\bar{N}\right)  \bar{k}%
^{\beta-1}\right)  n^{\ast}\left(  \bar{k},\bar{N}\right)  +A\alpha
\frac{\partial e^{\ast\alpha-1}}{\partial\bar{N}}\bar{k}^{\beta}\frac{\partial
n^{\ast}}{\partial\bar{k}}\bar{N}\\
&  =1-A\left(  \alpha\frac{\partial e^{\ast\alpha-1}}{\partial\bar{k}}\bar
{k}^{\beta}+\frac{\beta}{A}\right)  \frac{\partial n^{\ast}}{\partial\bar{N}%
}\bar{N}-A\left(  \alpha\frac{\partial e^{\ast\alpha-1}}{\partial\bar{k}}%
\bar{k}^{\beta}+\frac{\beta}{A}\right)  +A\alpha\frac{\partial e^{\ast
\alpha-1}}{\partial\bar{N}}\bar{k}^{\beta}\frac{\partial n^{\ast}}%
{\partial\bar{k}}\bar{N}>0
\end{align*}%
\begin{align*}
1-A\left(  \alpha\frac{\partial e^{\ast\alpha-1}}{\partial\bar{k}}\bar
{k}^{\beta}+\frac{\beta}{A}\right)  \frac{\partial n^{\ast}}{\partial\bar{N}%
}\bar{N} &  >A\left(  \alpha\frac{\partial e^{\ast\alpha-1}}{\partial\bar{k}%
}\bar{k}^{\beta}+\frac{\beta}{A}\right)  -A\alpha\frac{\partial e^{\ast
\alpha-1}}{\partial\bar{N}}\bar{k}^{\beta}\frac{\partial n^{\ast}}%
{\partial\bar{k}}\bar{N}\\
-A\left(  \alpha\frac{\partial e^{\ast\alpha-1}}{\partial\bar{k}}\bar
{k}^{\beta}+\frac{\beta}{A}\right)  \frac{\partial n^{\ast}}{\partial\bar{N}%
}\bar{N} &  >A\left(  \alpha\frac{\partial e^{\ast\alpha-1}}{\partial\bar{k}%
}\bar{k}^{\beta}+\frac{\beta}{A}\right)  -A\alpha\frac{\partial e^{\ast
\alpha-1}}{\partial\bar{N}}\bar{k}^{\beta}\frac{\partial n^{\ast}}%
{\partial\bar{k}}\bar{N}-1\\
-\frac{\partial n^{\ast}}{\partial\bar{N}}\bar{N} &  >\frac{A\left(
\alpha\frac{\partial e^{\ast\alpha-1}}{\partial\bar{k}}\bar{k}^{\beta}%
+\frac{\beta}{A}\right)  -A\alpha\frac{\partial e^{\ast\alpha-1}}{\partial
\bar{N}}\bar{k}^{\beta}\frac{\partial n^{\ast}}{\partial\bar{k}}\bar{N}%
-1}{A\left(  \alpha\frac{\partial e^{\ast\alpha-1}}{\partial\bar{k}}\bar
{k}^{\beta}+\frac{\beta}{A}\right)  }\\
-\frac{\partial n^{\ast}}{\partial\bar{N}}\bar{N} &  >1-\frac{\alpha
\frac{\partial e^{\ast\alpha-1}}{\partial\bar{N}}\bar{k}^{\beta}\frac{\partial
n^{\ast}}{\partial\bar{k}}\bar{N}}{\alpha\frac{\partial e^{\ast\alpha-1}%
}{\partial\bar{k}}\bar{k}^{\beta}+\frac{\beta}{A}}-\frac{1}{A\left(
\alpha\frac{\partial e^{\ast\alpha-1}}{\partial\bar{k}}\bar{k}^{\beta}%
+\frac{\beta}{A}\right)  }%
\end{align*}
The sufficient conditions for (i)-(iii) to be satisfied are:%
\begin{align}
1-\frac{\alpha\frac{\partial e^{\ast\alpha-1}}{\partial\bar{N}}\bar{k}^{\beta
}\frac{\partial n^{\ast}}{\partial\bar{k}}\bar{N}}{\alpha\frac{\partial
e^{\ast\alpha-1}}{\partial\bar{k}}\bar{k}^{\beta}+\frac{\beta}{A}}-\frac
{1}{A\alpha\frac{\partial e^{\ast\alpha-1}}{\partial\bar{k}}\bar{k}^{\beta
}+\beta} &  <-\frac{\partial n^{\ast}}{\partial\bar{N}}\bar{N}<1-\frac
{A\alpha\frac{\partial e^{\ast\alpha-1}}{\partial\bar{N}}\bar{k}^{\beta}%
\frac{\partial n^{\ast}}{\partial\bar{k}}\bar{N}}{1+A\alpha\frac{\partial
e^{\ast\alpha-1}}{\partial\bar{k}}\bar{k}^{\beta}+\beta} \nonumber \\ \label{stab_subs_appendix} \\
A\alpha\frac{\partial e^{\ast\alpha-1}}{\partial\bar{k}}\bar{k}^{\beta}+\beta
&  >1 \nonumber
\end{align}
where
\begin{align*}
    \frac{\partial n^{*}}{\partial \bar{k}} = -\frac{1-\alpha}{2 \bar{k}^{2}
\bar{N}^{\gamma}}, \qquad \qquad \frac{\partial n^{*}}{\partial \bar{N}} = -\frac{\gamma(1-\alpha)}{2 \bar{k}
\bar{N}^{\gamma+1}}, \\
\frac{\partial e^{*}}{\partial \bar{k}} = \frac{2\alpha}{1-\alpha} \, \bar{k}
\bar{N}^{\gamma}, \qquad \qquad \frac{\partial e^{*}}{\partial \bar{N}} = \frac{\alpha\gamma}{1-\alpha} \,
\bar{k}^{2} \bar{N}^{\gamma-1}
\end{align*}

\subsection{Proof of Proposition 3} 

When the time dad and mom dedicate to parenting are complements, from the
equilibrium the conditions, we obtain the following isoquants:%
\begin{align}
\bar{k} &  =A^{\frac{1}{1-\beta-2\alpha}}\left[  2\left(  \frac{\alpha
}{1-\alpha}\right)  \phi\bar{N}^{\gamma}\right]  ^{\frac{\alpha}%
{1-\beta-2\alpha}}\label{iso1_comp}\\
\bar{k} &  =\left(  \frac{1-\alpha}{4\phi\bar{N}^{\gamma}}\right)
\label{iso2_comp}%
\end{align}
Combining (\ref{iso1_comp}) and (\ref{iso2_comp}), we have:%
\[
\frac{1-\alpha}{4\phi\bar{N}^{\gamma}}=A^{\frac{1}{1-\beta-2\alpha}}\left[
2\left(  \frac{\alpha}{1-\alpha}\right)  \phi\bar{N}^{\gamma}\right]
^{\frac{\alpha}{1-\beta-2\alpha}}%
\]
Isolating $\bar{N}$, it follows:%
\begin{align*}
\frac{1-\alpha}{4\phi}  & =A^{\frac{1}{1-\beta-2\alpha}}\left[  2\left(
\frac{\alpha}{1-\alpha}\right)  \phi\right]  ^{\frac{\alpha}{1-\beta-2\alpha}%
}\bar{N}^{\gamma+\frac{\alpha\gamma}{1-\beta-2\alpha}}\\
\left(  \frac{1-\alpha}{4\phi}\right)  A^{\frac{-1}{1-\beta-2\alpha}}\left[
2\left(  \frac{\alpha}{1-\alpha}\right)  \phi\right]  ^{\frac{-\alpha}%
{1-\beta-2\alpha}}  & =\bar{N}^{\frac{\left(  1-\beta-\alpha\right)  \gamma
}{1-\beta-2\alpha}}\\
\bar{N}^{\left(  1-\beta-\alpha\right)  \gamma}  & =\left(  \frac{1-\alpha
}{4\phi}\right)  ^{1-\beta-2\alpha}A^{-1}\left[  2\left(  \frac{\alpha
}{1-\alpha}\right)  \phi\right]  ^{-\alpha}\\
\bar{N}  & =A^{\frac{-1}{\left(  1-\beta-\alpha\right)  \gamma}}\left(
\frac{1-\alpha}{4\phi}\right)  ^{\frac{1-\beta-2\alpha}{\left(  1-\beta
-\alpha\right)  \gamma}}\left[  2\left(  \frac{\alpha}{1-\alpha}\right)
\phi\right]  ^{\frac{-\alpha}{\left(  1-\beta-\alpha\right)  \gamma}}%
\end{align*}
Notice we can rewrite it as:%
\begin{align}
\bar{N}  & =2^{\frac{-1+\beta+2\alpha}{\left(  1-\beta-\alpha\right)  \gamma}%
}2^{\frac{-\alpha}{\left(  1-\beta-\alpha\right)  \gamma}}A^{\frac{-1}{\left(
1-\beta-\alpha\right)  \gamma}}\left(  \frac{1-\alpha}{2\phi}\right)
^{\frac{1-\beta-2\alpha}{\left(  1-\beta-\alpha\right)  \gamma}}\left[
\left(  \frac{\alpha}{1-\alpha}\right)  \phi\right]  ^{\frac{-\alpha}{\left(
1-\beta-\alpha\right)  \gamma}}\nonumber\\
& =2^{\frac{-1+\beta+\alpha}{\left(  1-\beta-\alpha\right)  \gamma}}%
A^{\frac{-1}{\left(  1-\beta-\alpha\right)  \gamma}}\left(  \frac{1-\alpha
}{2\phi}\right)  ^{\frac{1-\beta-2\alpha}{\left(  1-\beta-\alpha\right)
\gamma}}\left[  \left(  \frac{\alpha}{1-\alpha}\right)  \phi\right]
^{\frac{-\alpha}{\left(  1-\beta-\alpha\right)  \gamma}}\nonumber\\
& =\Omega A^{\frac{\gamma}{\left(  1-\beta\right)  \gamma+\left(
1-\gamma\right)  \alpha}}\underset{\text{{\scriptsize Fertility}}}%
{\underbrace{\left(  \frac{1-\alpha}{2\phi}\right)  }}^{\frac{\alpha\gamma
}{\left(  1-\beta-\alpha\right)  \gamma}}\underset
{\text{{\scriptsize Education}}}{\underbrace{\left[  \left(  \frac{\alpha
}{1-\alpha}\right)  \phi\right]  }}^{\frac{\alpha\gamma}{\left(
1-\beta-\alpha\right)  \gamma}}\label{eq_N_comp}%
\end{align}
where%
\[
\Omega=2^{\frac{-1+\beta+\alpha}{\left(  1-\beta-\alpha\right)  \gamma}%
}=2^{-\frac{1}{\gamma}}%
\]
Because $\gamma>0$, it follows that $0<\Omega<1$.

Finally, substituting (\ref{eq_N_comp}) into (\ref{iso2_comp}), we obtain:%
\begin{align*}
\bar{k}  & =\left(  \frac{1-\alpha}{4\phi}\right)  2^{\frac{-\left(
-1+\beta+\alpha\right)  \gamma}{\left(  1-\beta-\alpha\right)  \gamma}%
}A^{\frac{\gamma}{\left(  1-\beta-\alpha\right)  \gamma}}\left(
\frac{1-\alpha}{2\phi}\right)  ^{\frac{-\left(  1-\beta-2\alpha\right)
\gamma}{\left(  1-\beta-\alpha\right)  \gamma}}\left[  \left(  \frac{\alpha
}{1-\alpha}\right)  \phi\right]  ^{\frac{\alpha\gamma}{\left(  1-\beta
-\alpha\right)  \gamma}}\\
& =\left(  \frac{1-\alpha}{4\phi}\right)  2A^{\frac{1}{1-\beta-\alpha}}\left(
\frac{1-\alpha}{2\phi}\right)  ^{\frac{-1+\beta+2\alpha}{1-\beta-\alpha}%
}\left[  \left(  \frac{\alpha}{1-\alpha}\right)  \phi\right]  ^{\frac{\alpha
}{1-\beta-\alpha}}\\
& =A^{\frac{1}{1-\beta-\alpha}}\left(  \frac{1-\alpha}{2\phi}\right)
^{\frac{\alpha}{1-\beta-\alpha}}\left[  \left(  \frac{\alpha}{1-\alpha
}\right)  \phi\right]  ^{\frac{\alpha}{1-\beta-\alpha}}\\
& =A^{\frac{1}{1-\beta-\alpha}}\left(  \frac{\alpha}{2}\right)  ^{\frac
{\alpha}{1-\beta-\alpha}}\\
& =\left(  A^{\frac{1}{\alpha}}\frac{\alpha}{2}\right)  ^{\frac{\alpha
}{1-\beta-\alpha}}%
\end{align*}

\subsection{Proof of Proposition 4} 

Given that our dynamic system can still be specified as in (\ref{Dynam_Sys_base}), the local stability conditions for the non-trivial equilibrium point remain the same as in (\ref{stab_subs_appendix}). But now, through direct derivation of the optimal allocations, we have:
\begin{align*}
    \frac{\partial n^*}{\partial \bar{k}}
= -\frac{1-\alpha}{4 \bar{k}^2 \bar{N}^{\gamma}}, \qquad \qquad \frac{\partial n^*}{\partial \bar{N}}
= -\frac{\gamma (1-\alpha)}{4 \bar{k} \bar{N}^{\gamma+1}}, \\
\frac{\partial e^*}{\partial \bar{k}}
= \frac{4\alpha}{1-\alpha} \bar{k} \bar{N}^{\gamma}, \qquad \qquad \frac{\partial e^*}{\partial \bar{N}}
= \frac{2\alpha\gamma}{1-\alpha} \bar{k}^2 \bar{N}^{\gamma-1}
\end{align*}

\subsection{Proof of Proposition 5}

When the time dad and mom dedicate to parenting are substitutes, from the
equilibrium conditions, and recalling that at a non-trivial steady state
$\bar{n}=1$, we obtain the following isoquants:
\begin{align}
\bar{k}
&=
A^{\frac{1}{1-\beta-\alpha}}
\left[
\frac{\alpha}{1+\alpha}
\left(
1-\bar{k}\phi \bar{N}^{\gamma}
\right)
\right]^{\frac{\alpha}{1-\beta-\alpha}}
\label{iso1_subs_b}\\
\bar{k}
&=
\frac{(1+\alpha)b}
{\left(
1-\bar{k}\phi \bar{N}^{\gamma}
\right)
\left[
1+\alpha-2\left(
1-\bar{k}\phi \bar{N}^{\gamma}
\right)
\right]}
\label{iso2_subs_b}
\end{align}

The first expression follows from the human-capital law of motion together with
the fact that, at the steady state, $\bar{c}+\bar{e}=\bar{y}$ and
$\bar{e}=\alpha\bar{c}$, so that:
\[
\bar{e}=\frac{\alpha}{1+\alpha}\bar{y}.
\]
Under perfect substitution, output is:
\[
\bar{y}
=
\bar{k}-\bar{k}^{2}\phi\bar{N}^{\gamma}
=
\bar{k}\left(1-\bar{k}\phi\bar{N}^{\gamma}\right),
\]
which yields (\ref{iso1_subs_b}).

The second expression follows from the corrected quadratic condition
(\ref{n_quad_subs}) evaluated at $\bar{n}=1$. Using
$X=\bar{k}^{2}\phi\bar{N}^{\gamma}$, we have:
\[
2X^{2}
+
\bar{k}\left[(\alpha-3)X+(1+\alpha)b\right]
+
(1-\alpha)\bar{k}^{2}
=
0.
\]
Substituting back for $X$ and dividing through by $\bar{k}^{2}>0$, it follows
that:
\[
2\bar{k}^{2}\phi^{2}\bar{N}^{2\gamma}
+
(\alpha-3)\bar{k}\phi\bar{N}^{\gamma}
+
(1+\alpha)\frac{b}{\bar{k}}
+
(1-\alpha)
=
0.
\]
Rearranging:
\[
\left(
1-\bar{k}\phi\bar{N}^{\gamma}
\right)
\left[
1+\alpha-2\left(
1-\bar{k}\phi\bar{N}^{\gamma}
\right)
\right]
=
(1+\alpha)\frac{b}{\bar{k}},
\]
from which (\ref{iso2_subs_b}) follows.

To combine the two equilibrium conditions, define
\[
z\equiv 1-\bar{k}\phi\bar{N}^{\gamma}.
\]
Then (\ref{iso1_subs_b}) and (\ref{iso2_subs_b}) become:
\begin{align}
\bar{k}
&=
A^{\frac{1}{1-\beta-\alpha}}
\left(
\frac{\alpha z}{1+\alpha}
\right)^{\frac{\alpha}{1-\beta-\alpha}}
\label{iso1_subs_b_z}\\
\bar{k}
&=
\frac{(1+\alpha)b}{z(1+\alpha-2z)}.
\label{iso2_subs_b_z}
\end{align}

Combining (\ref{iso1_subs_b_z}) and (\ref{iso2_subs_b_z}), we have:
\[
A^{\frac{1}{1-\beta-\alpha}}
\left(
\frac{\alpha z}{1+\alpha}
\right)^{\frac{\alpha}{1-\beta-\alpha}}
=
\frac{(1+\alpha)b}{z(1+\alpha-2z)}.
\]
Rearranging:
\begin{equation}
A^{\frac{1}{1-\beta-\alpha}}
\left(
\frac{\alpha}{1+\alpha}
\right)^{\frac{\alpha}{1-\beta-\alpha}}
z^{\frac{1-\beta}{1-\beta-\alpha}}
(1+\alpha-2z)
=
(1+\alpha)b.
\label{eq_z_subs_b}
\end{equation}

Thus, feasible steady states are associated with the roots $z_i$,
$i=1,2$, of (\ref{eq_z_subs_b}) over the interval
$z\in\left(0,\frac{1+\alpha}{2}\right)$.

To study the number of solutions, notice that the left-hand side of
(\ref{eq_z_subs_b}) is equal to zero at $z=0$ and at
$z=\frac{1+\alpha}{2}$. Since $0<\alpha+\beta<1$, the exponent
\[
\frac{1-\beta}{1-\beta-\alpha}>1,
\]
and differentiating the left-hand side of (\ref{eq_z_subs_b}) with respect to
$z$ gives:
\[
A^{\frac{1}{1-\beta-\alpha}}
\left(
\frac{\alpha}{1+\alpha}
\right)^{\frac{\alpha}{1-\beta-\alpha}}
z^{\frac{\alpha}{1-\beta-\alpha}}
\left[
\frac{1-\beta}{1-\beta-\alpha}(1+\alpha)
-
2\left(
\frac{1-\beta}{1-\beta-\alpha}+1
\right)z
\right].
\]
Hence, the left-hand side of (\ref{eq_z_subs_b}) is single-peaked over
$\left(0,\frac{1+\alpha}{2}\right)$. Therefore, there exists a threshold value
$\bar{b}>0$ such that (\ref{eq_z_subs_b}) admits two feasible roots if
$0<b<\bar{b}$, one double root if $b=\bar{b}$, and no feasible root if
$b>\bar{b}$.

Finally, once a feasible root $z_i$ is known, it follows from
(\ref{iso1_subs_b_z}) that:
\[
\bar{k}_i^b
=
A^{\frac{1}{1-\beta-\alpha}}
\left(
\frac{\alpha z_i}{1+\alpha}
\right)^{\frac{\alpha}{1-\beta-\alpha}},
\]
and since
\[
\bar{k}_i^b\phi(\bar{N}_i^b)^\gamma=1-z_i,
\]
we obtain:
\[
\bar{N}_i^b
=
\left(
\frac{1-z_i}{\phi\bar{k}_i^b}
\right)^{\frac{1}{\gamma}}.
\]

\subsection{Proof of Proposition 6}

The proof follows the same steps as in Appendix A.2. For each optimal
allocation $(n_{Si}^*,e_{Si}^*)$, $i=1,2$, the distorted dynamic system can
still be written in the general form:
\begin{align}
k_{t+1}  &  =Ae_i^{\ast\alpha}\left(  k_{t},N_{t}\right)  k_{t}^{\beta
}\nonumber\\
& \label{Dynam_Sys_base_b}\\
N_{t+1}  &  =n_i^{\ast}\left(  k_{t},N_{t}\right)  N_{t}\nonumber
\end{align}
where $e_i^{\ast}\left(  k_{t},N_{t}\right)$ and $n_i^{\ast}\left(  k_{t},N_{t}\right)$
are the distorted education investment and fertility rates associated with the
feasible root $n_{Si}^*$ of (\ref{n_quad_subs}).

At a non-trivial steady state,
$k_{t}=k_{t+1}=\bar{k}$ and $N_{t}=N_{t+1}=\bar{N}$, so it follows that:
\begin{align}
e_i^{\ast\alpha}\left(  \bar{k},\bar{N}\right)
&=\frac{\bar{k}^{1-\beta}}{A}\nonumber\\
& \label{Auxiliary_b}\\
n_i^{\ast}\left(  \bar{k},\bar{N}\right)
&=1\nonumber
\end{align}

Thus, the Jacobian matrix and the local stability conditions are exactly the
same as in Appendix A.2. Therefore, each equilibrium point
$P_{Si}=(\bar{k}_{Si},\bar{N}_{Si})$ is locally stable provided that
(\ref{stab_subs_b}) is satisfied.

The only difference relative to Proposition 2 is that fertility and education
are now determined by distorted policy functions. In particular, distorted
fertility is implicitly defined by:
\[
F_S(n,k,N)
=
2X^2n^2+k\left[(\alpha-3)X+(1+\alpha)b\right]n+(1-\alpha)k^2=0,
\]
where $X=k^2\phi N^\gamma$. Applying the implicit function theorem:
\[
\frac{\partial n_i^\ast}{\partial k}
=
-\frac{F_{S,k}}{F_{S,n}},
\qquad \qquad
\frac{\partial n_i^\ast}{\partial N}
=
-\frac{F_{S,N}}{F_{S,n}},
\]
where
\begin{align*}
F_{S,n}
&=
4X^2n_i^\ast+k\left[(\alpha-3)X+(1+\alpha)b\right],\\
F_{S,k}
&=
\frac{8X^2 {n_i^\ast}^2}{k}+\left[3(\alpha-3)X+(1+\alpha)b\right]n_i^\ast+2(1-\alpha)k,\\
F_{S,N}
&=
\frac{\gamma X n_i^\ast}{N}\left[4Xn_i^\ast+(\alpha-3)k\right].
\end{align*}

The distorted education rule can be written as:
\[
e_i^\ast
=
\frac{\alpha}{1-\alpha}G_S(k,N),
\]
where
\[
G_S(k,N)
=
\frac{k(X-b)-X^2n_i^\ast}{k-Xn_i^\ast}.
\]
Therefore:
\[
\frac{\partial e_i^\ast}{\partial k}
=
\frac{\alpha}{1-\alpha}G_{S,k},
\qquad \qquad
\frac{\partial e_i^\ast}{\partial N}
=
\frac{\alpha}{1-\alpha}G_{S,N},
\]
where
\begin{align*}
G_{S,k}
&=
\frac{
\left[
3X-b-\frac{4X^2n_i^\ast}{k}-X^2\frac{\partial n_i^\ast}{\partial k}
\right]
(k-Xn_i^\ast)
-
\left[
k(X-b)-X^2n_i^\ast
\right]
\left[
1-\frac{2Xn_i^\ast}{k}-X\frac{\partial n_i^\ast}{\partial k}
\right]
}{
(k-Xn_i^\ast)^2
},\\[1em]
G_{S,N}
&=
\frac{
\left[
\frac{\gamma X}{N}(k-2Xn_i^\ast)-X^2\frac{\partial n_i^\ast}{\partial N}
\right]
(k-Xn_i^\ast)
+
\left[
k(X-b)-X^2n_i^\ast
\right]
\left[
\frac{\gamma X n_i^\ast}{N}+X\frac{\partial n_i^\ast}{\partial N}
\right]
}{
(k-Xn_i^\ast)^2
}.
\end{align*}

Finally, making use of the chain rule:
\[
\frac{\partial e_i^{\ast\alpha}}{\partial k}
=
\alpha e_i^{\ast\alpha-1}\frac{\partial e_i^\ast}{\partial k},
\qquad \qquad
\frac{\partial e_i^{\ast\alpha}}{\partial N}
=
\alpha e_i^{\ast\alpha-1}\frac{\partial e_i^\ast}{\partial N}.
\]
Substituting these distorted policy derivatives into the same Jacobian
expressions used in Appendix A.2 yields (\ref{stab_subs_b}). 

\subsection{Proof of Proposition 7}

The proof follows the same logic as in Appendix A.5, with the same shortcut
used in Appendix A.3. When the time dad and mom dedicate to parenting are complements, from the equilibrium conditions, and recalling that at a non-trivial steady state $\bar{n}=1$, we obtain the following isoquants:
\begin{align}
\bar{k}
&=
A^{\frac{1}{1-\beta-\alpha}}
\left[
\frac{\alpha}{1+\alpha}
\left(
1-2\bar{k}\phi \bar{N}^{\gamma}
\right)
\right]^{\frac{\alpha}{1-\beta-\alpha}}
\label{iso1_comp_b}\\
\bar{k}
&=
\frac{(1+\alpha)b}
{\left(
1-2\bar{k}\phi \bar{N}^{\gamma}
\right)
\left[
1+\alpha-2\left(
1-2\bar{k}\phi \bar{N}^{\gamma}
\right)
\right]}
\label{iso2_comp_b}
\end{align}
Now define
\[
z\equiv 1-2\bar{k}\phi \bar{N}^{\gamma}.
\]
Then (\ref{iso1_comp_b}) and (\ref{iso2_comp_b}) become:
\begin{align}
\bar{k}
&=
A^{\frac{1}{1-\beta-\alpha}}
\left(
\frac{\alpha z}{1+\alpha}
\right)^{\frac{\alpha}{1-\beta-\alpha}}
\label{iso1_comp_b_z}\\
\bar{k}
&=
\frac{(1+\alpha)b}{z(1+\alpha-2z)}.
\label{iso2_comp_b_z}
\end{align}

Thus, the scalar equation determining feasible steady states is exactly the
same as in (\ref{eq_z_subs_b}). It follows immediately that the feasible roots
$z_i$, $i=1,2$, are the same as in Proposition \ref{prop 5}, and therefore:
\[
\bar{k}_{Ci}=\bar{k}_{Si}, \qquad i=1,2.
\]

The only difference is in the population term. Under perfect substitution:
\[
\bar{k}_{Si}\phi\bar{N}_{Si}^{\gamma}=1-z_i,
\]
whereas under perfect complementarity:
\[
2\bar{k}_{Ci}\phi\bar{N}_{Ci}^{\gamma}=1-z_i,
\]
or equivalently:
\[
\bar{k}_{Ci}\phi\bar{N}_{Ci}^{\gamma}=\frac{1-z_i}{2}.
\]
Since $\bar{k}_{Ci}=\bar{k}_{Si}$, it follows that:
\[
\bar{N}_{Ci}^{\gamma}
=
\frac{1}{2}\bar{N}_{Si}^{\gamma},
\]
and therefore:
\[
\bar{N}_{Ci}
=
\frac{1}{2^{1/\gamma}}\bar{N}_{Si}
=
\Omega\bar{N}_{Si}.
\]

\subsection{Proof of Proposition 8}

The proof is identical in structure to Appendix A.6 and closely parallels
Appendix A.4. For each optimal allocation $(n_{Ci}^*,e_{Ci}^*)$, $i=1,2$, the
distorted dynamic system can still be written as in (\ref{Dynam_Sys_base_b}).
Therefore, the Jacobian matrix and the local stability conditions remain the
same as in Appendix A.2. Hence, each equilibrium point
$P_{Ci}=(\bar{k}_{Ci},\bar{N}_{Ci})$ is locally stable provided that
(\ref{stab_subs_b}) is satisfied.

The only difference is that fertility and education are now determined by
distorted complementary policy functions. In particular, distorted fertility is
implicitly defined by:
\[
F_C(n,k,N)
=
8X^2n^2+k\left[2(\alpha-3)X+(1+\alpha)b\right]n+(1-\alpha)k^2=0,
\]
where $X=k^2\phi N^\gamma$. Applying the implicit function theorem:
\[
\frac{\partial n_i^\ast}{\partial k}
=
-\frac{F_{C,k}}{F_{C,n}},
\qquad \qquad
\frac{\partial n_i^\ast}{\partial N}
=
-\frac{F_{C,N}}{F_{C,n}},
\]
where
\begin{align*}
F_{C,n}
&=
16X^2n_i^\ast+k\left[2(\alpha-3)X+(1+\alpha)b\right],\\
F_{C,k}
&=
\frac{32X^2{n_i^\ast}^2}{k}+\left[6(\alpha-3)X+(1+\alpha)b\right]n_i^\ast+2(1-\alpha)k,\\
F_{C,N}
&=
\frac{2\gamma X n_i^\ast}{N}\left[8Xn_i^\ast+(\alpha-3)k\right].
\end{align*}

The distorted complementary education rule can be written as:
\[
e_i^\ast
=
\frac{\alpha}{1-\alpha}G_C(k,N),
\]
where
\[
G_C(k,N)
=
\frac{k(2X-b)-4X^2n_i^\ast}{k-2Xn_i^\ast}.
\]
Therefore:
\[
\frac{\partial e_i^\ast}{\partial k}
=
\frac{\alpha}{1-\alpha}G_{C,k},
\qquad \qquad
\frac{\partial e_i^\ast}{\partial N}
=
\frac{\alpha}{1-\alpha}G_{C,N},
\]
where
\begin{align*}
G_{C,k}
&=
\frac{
\left[
6X-b-\frac{16X^2n_i^\ast}{k}-4X^2\frac{\partial n_i^\ast}{\partial k}
\right]
(k-2Xn_i^\ast)
-
\left[
k(2X-b)-4X^2n_i^\ast
\right]
\left[
1-\frac{4Xn_i^\ast}{k}-2X\frac{\partial n_i^\ast}{\partial k}
\right]
}{
(k-2Xn_i^\ast)^2
},\\[1em]
G_{C,N}
&=
\frac{
\left[
\frac{2\gamma X}{N}(k-4Xn_i^\ast)-4X^2\frac{\partial n_i^\ast}{\partial N}
\right]
(k-2Xn_i^\ast)
+
\left[
k(2X-b)-4X^2n_i^\ast
\right]
\left[
\frac{2\gamma X n_i^\ast}{N}+2X\frac{\partial n_i^\ast}{\partial N}
\right]
}{
(k-2Xn_i^\ast)^2
}.
\end{align*}

Finally, making use of the chain rule:
\[
\frac{\partial e_i^{\ast\alpha}}{\partial k}
=
\alpha e_i^{\ast\alpha-1}\frac{\partial e_i^\ast}{\partial k},
\qquad \qquad
\frac{\partial e_i^{\ast\alpha}}{\partial N}
=
\alpha e_i^{\ast\alpha-1}\frac{\partial e_i^\ast}{\partial N}.
\]
Substituting these distorted policy derivatives into the same Jacobian expressions used in Appendix A.2 yields the result.

\newpage

\setlength{\bibsep}{4pt}
\bibliography{reference}

\end{document}